\documentclass{JHEP3}
\usepackage{amsfonts}
\usepackage{amsmath,amssymb,epsfig}

\preprint{Napoli DSF-T-10/2004\\INFN-NA-10/2004\\ISAS/40/2004/FM}
\begin{abstract}
{ We analyze the modular properties of the effective CFT description for
Jain plateaux, proposed in \cite{cgm1}, corresponding to the fillings
$\nu=\frac{ m}{2pm+1}$. We construct its characters for the twisted and the
untwisted sector and the diagonal partition function. We show that the
degrees of freedom entering the partition function go to complete a
$Z_{m}$-orbifold construction of the RCFT $\widehat{u(1)}{{\times}
}\widehat{su(m)}_{1}$ proposed for the Jain states \cite{Frohlich,
Cappelli}. The resulting extended algebra of the chiral primary fields can
be also viewed as a RCFT extension of the $\widehat{u(1)}{{\times} }{\cal W}_{m}$
minimal models \cite{Cappelli}. For $m=2$ we prove that our model, the TM,
gives the RCFT closure of the extended minimal models $\widehat{u(1)} {{\times}
}{\cal W}_{2}$.}
\end{abstract}

\title{Jain states on a torus: an unifying description\thanks{
Work supported in part by the European Communities Human Potential Program
under contract HPRN-CT-2000-00131 Quantum Spacetime}}
\author{ Gerardo Cristofano, Vincenzo Marotta \\ Dipartimento di
Scienze Fisiche, Universit\'{a} di Napoli ``Federico II'' and INFN, Sezione di
Napoli - Via Cintia - Compl. universitario M. Sant'Angelo
- 80126 Napoli, Italy }
\author { Giuliano Niccoli \\ Sissa and INFN, Sezione di
Trieste - Via Beirut 1 - 34100 Trieste, Italy}

\begin{document}

\section{Introduction}

Historically the motivations for the use of the Effective Conformal Field
Theory (ECFT) in describing a Quantum Hall Fluid (QHF) at the plateau
\cite{wen, Stone} go back to the observation that the ground state wave
function (Laughlin states) for the filling $\nu =1/(2p+1)$ \cite{laugh} can
be written as correlators of vertex operators describing the anyon states
(charged states), which were identified as primary states of a two
dimensional chiral CFT with central charge $c=1$ \cite{lufub}. The ECFT
description of a QHF at the plateau accounts for the incompressibility of
the Laughlin fluid. The dynamical symmetry of this fluid in the disk
geometry is the area-preserving diffeomorphisms of the plane which imply
the $W_{1+\infty }$ algebra. This is the unique centrally extended quantum
analogue of the classical area preserving diffeomorphisms algebra
$w_{\infty }$ on the circle. For general filling it is known that it is
necessary to introduce neutral fields in order to describe the
area-preserving edge deformations of the incompressible fluid. In this
context two classes of CFTs have been proposed for Jain plateaux: the
multi-components theory \cite{Frohlich}, characterized by the extended
algebra $\widehat{u(1)}{{\times} }\widehat{ su(m)}_{1}$ and the $W_{1+\infty }$
minimal models \cite{ctz5} both with central charge $c=m$. In spite of
having the same spectrum of edge excitations, they manifest differences in
the degeneracy of the states and in the quantum statistics.

An alternative and simple construction of the hierarchical model based upon
the so called $m$-reduction procedure was given in \cite{cgm1}, where a
$c=m$ (daughter) CFT was obtained from a $c=1$ (mother) one. The daughter
theory preserves the $W_{1+\infty }$ symmetry of the mother theory. In fact
it is well known that for any positive integer $m$, the $W_{1+\infty }$
algebra with central charge $c$ is isomorphic to its subalgebra consisting
of elements of degrees divisible by $m$ and which appear with central
charge $ mc $ \cite{FKN, bakalov}, so implying the incompressibility of the
QHF for any given plateau.

This procedure was developed in \cite{VM} and employed for both the Jain
fillings $\nu=\frac{m}{2pm+1}$ in \cite{cgm1} and the non standard ones
$\nu=\frac{m}{pm+2}$ in \cite{cgm2} where it was shown how naturally it
induces twists on the boundary, given by a $Z_{m}$ group generated by the
phases $\epsilon ^{j}=e^{\frac{2\pi ij}{m }}$ , $j=1,...,m$. As a
consequence the chiral primary fields naturally appear as composite
operators with a charged and a neutral component. It has been proved that
their correlators reproduce the ground state wave function for the Jain as
well as for the paired states, so generalizing the Laughlin type wave
function on the plane \cite{cgm1, cgm2}. Only one $\widehat{u(1)}$ charged
current survives to the discrete twist group while the role of the neutral
degrees of freedom is to cancel out the $m$-root singularity present in the
correlators of the charged operators, so restoring locality.

In this paper we present a detailed analysis of such a model, the Twisted
Model (TM), for the Jain states on a torus (genus $g=1$ Riemann surface),
so extending the previous description given on the plane \cite{cgm1}. For
the paired states this analysis was given in \cite{Paired2}.

We construct the characters associated to the extended chiral algebra and
show that they provide a finite dimensional representation of the $%
\Gamma _{0}(2)$ subgroup of the modular group $SL(2,Z)/Z_{2}$. Finally we
exhibit the diagonal partition function which is invariant under such a
subgroup. The resulting theory is the $Z_{m}$-orbifold of the Rational CFT
(RCFT) describing the multi-components theory. Furthermore the extended
algebra of the TM can be also viewed as a particular RCFT extension of the
$ W_{1+\infty }$ minimal models\footnote{We recall that the $W_{1+\infty }$
minimal models \cite{KT}, which are defined by the full degenerate
representations of $W_{1+\infty },$ are not a RCFT because it is impossible
to build a finite extended algebra by assembling their primary fields, each
taken with multiplicity one, that is closed under the S modular
transformation. Instead the multi-components bosonic theories are RCFT
extensions of the $W_{1+\infty }$ minimal models \cite{ctz5, KT, marina}
where the extended primary fields are defined by collecting the fields of
the full degenerate $W_{1+\infty }$ representations, but with multiplicity
different from one. The TM is also a RCFT extension of the theory with a $
W_{1+\infty }$ symmetry in which there are different multiplicities in any
different sector.} \cite{KT}.

Indeed their extensions are the constituents of the periodic-periodic (\
P-P) and periodic-antiperiodic (P-A) sector of this orbifold construction
respectively. In addition to them the twisted (A-P) sector, is exactly what
is needed in order to obtain the closure under $\Gamma _{0}(2)$. We should
also comment on the presence of quasihole states in the twisted sector and
clustering properties \cite{cgm1, PS} as it happens in the paired states.

All that emphasizes the key role of the $m$-reduction procedure in producing
a RCFT which realizes a consistent $W_{1+\infty }$\ extension of the
classical area-preserving diffeomorphisms of the plane.

The paper is organized as follows:

In sec. \ref{kmatrix} we review the K-matrix construction \cite{Frohlich}
of the multi-components theory of the Jain fillings and briefly summarize
the properties of the ground state wave functions on the plane of a system
of interacting layers which have been first analyzed in \cite{Halperin0}.

In sec. \ref{symmetry} we review the symmetries of the ECFT which are known
in the literature: the RCFT with symmetry $\widehat{u(1)}{{\ {\times}} }%
\widehat{su(m)}_{1}$ and the $W_{1+\infty }$ minimal models \cite{ctz5}. We
also give the decomposition of the $\widehat{u(1)}{{\times} }\widehat{su(m)}%
_{1}$ characters in terms of those of the $W_{1+\infty }$ minimal models
\cite{ctz5, KT, marina}. Finally we indicate the $Z_{m}$ discrete group
with respect to which the orbifold construction is made.

In sec. \ref{planereduction} we briefly summarize the $m$-reduction
construction on the plane (genus $g=0$), which has been used in \cite{cgm1}
to construct the TM for the Jain series at filling $\nu =\frac{m}{2pm+1}$.

In sec. \ref{torusreduction} we generalize it to the torus topology (genus
$ g=1$) and derive the extended chiral primary fields both for the twisted
and the untwisted sectors and the corresponding characters. Then a
comparison of the TM with the multi-components theory \cite{Frohlich} is
given and the features of the $Z_{m}$-orbifold construction shown. The
factorization of the characters in terms of charged and neutral components
is derived. Finally the special $m=2$ case is also analyzed.

In sec. \ref{partitionfunction} we construct the complete diagonal partition
function of the TM.

In sec. \ref{discussion} we show that the TM provides a RCFT closure of the
extended minimal models $\widehat{u(1)}{{\times} }\mathcal{W}_{m}{{{\times} } }Z_2$. In
fact we prove that the P-A sector of the TM gives a realization of the
chiral algebra containing only degenerate $W_{1+\infty }$ representations.

Finally in sec. \ref{conclusions} we give conclusions and comments.

In App. A we describe the properties of the $\Gamma _{0}(2)$ subgroup of
the modular group.

In App. B we describe the extended odd RCFT $\widehat{u(1)}%
_{q}$ for the Laughlin filling, $\nu =\frac{1}{q}$ where $q$ is odd, which
we use as mother theory for the $m$-reduction, pointing out the reason for
requesting the closure under $\Gamma _{0}(2)$ only.

In App. C we give the transformations of the characters of the TM under the
elements of $\Gamma _{0}(2)$, showing that they provide a finite
representation for it. We also prove that the TM diagonal partition function
is invariant under its action.

\section{K-matrix construction of the multi-components theory\label{kmatrix}}

The CFT for a $m$-levels QHF at filling $\nu =\frac{m}{2pm+1}$ was given in
\cite{Frohlich, Wen0, Wn} introducing the symmetric and positive $\mathbf{K}$
matrix:
\begin{equation}
\mathbf{K}=\mathbf{1}_{m{{\times} }m}+2p\mathbf{C}_{m{{\times} }%
m}\,\,\,\,where:\mathbf{C}_{m{{\times} }m}=\left\Vert 1\right\Vert _{m{{%
{\times}} }m}
\end{equation}%
The entries $\mathbf{K}_{\left( i,j\right) },$ depending on $m$ and $p$,
define the braiding factors between electrons in the $i$ and $j$ level. In
this simplified picture the levels are considered independent and fully
equivalent and we will refer to them also as layers. The wave function on
the plane for such a system can be given in terms of $\mathbf{K}$ as:
\begin{equation}
f^{MC}(\{z_{i}^{(a)}\})=\prod_{h=1}^{m}\prod_{i<j}\left(
z_{i}^{(h)}-z_{j}^{(h)}\right) ^{\mathbf{K}_{\left( h,h\right)
}}\prod_{1=h<k}^{m}\prod_{i<j}\left( z_{i}^{(h)}-z_{j}^{(k)}\right) ^{%
\mathbf{K}_{\left( h,k\right) }}e^{-\frac{1}{4}\sum_{h=1}^{m}%
\sum_{i}|z_{i}^{(h)}|^{2}}
\end{equation}%
This is a special class of more general trial functions for the ground state
of a multi-layer system. For the simplest case of only two layers (or
levels) they can be codified in the so called ($m$,$n$,$k$) states
introduced by Halperin \cite{Halperin0}. The integers in the bracket are in
correspondence with the entries of the $\mathbf{K}$ matrix, i.e. $m=\mathbf{%
K }_{\left( 1,1\right) }$; $n=\mathbf{K}_{\left( 2,2\right) }$ and $k=%
\mathbf{K}_{\left( 1,2\right) }$. Some values of these integers give
interesting models for special plateaux. The ($p+2$, $p+2$, $p$) generalize
the well studied ($3$,$3$,$1$) state and it was analyzed recently in the
m-reduction approach in \cite{Paired2}. Another interesting case is the
($1$,$1$,$1$) for which theoretical studies predict the existence of a
Josephson like effect. Nevertheless this state does not satisfy the
non-degeneracy of the $\mathbf{K}$ matrix so it does not enter the theories
we are studying now.

The CFT of the system with a general $\mathbf{K}$ matrix is realized by $m$
free bosons with an extended chiral algebra. These fields satisfy the
compactification conditions given by \cite{Frohlich}:
\begin{equation}
\mathbf{Q}(z)=\mathbf{Q}(z)+2\pi \mathbf{Rh}\,\,\,\,\,\,\,
\label{compattificazione}
\end{equation}%
where $\,\,\mathbf{h}\in \mathbb{Z}^{m}$ and $\mathbf{Q}(z)$ is the vector
field with bosonic components $Q^{(i)}(z),$ $i=1,...,m.$

The $m{{\times} }m$ matrix $\mathbf{R}$ is defined by\footnote{$\mathbf{R}$\
is the positive root of $\mathbf{K}$ and is well defined because $\mathbf{K}$
is a symmetric and positive definite matrix.}:
\begin{equation}
\mathbf{R}^{T}\mathbf{R}=\mathbf{K}
\end{equation}%
and it is given explicitly by
\begin{equation}
\mathbf{R}=\mathbf{1}_{m{{\times} }m}+\frac{1}{m}\left( \sqrt{2pm+1}%
-1\right) \mathbf{C}_{m{\ {{\times} }}m}.
\end{equation}%
It is actually a metric of the lattice $\Gamma $ which refers to the
electron field. The compactification matrix (\ref{compattificazione}) for
diagonal $\mathbf{h}=h\mathbf{t}$, where $\mathbf{t}$ is the charge vector
which satisfies the condition $\mathbf{t}^{T}\left( \mathbf{K}\right) ^{-1}%
\mathbf{t}=\nu $ and coincides with $\mathbf{t}^{T}=\left( 1,..,1\right) $,
defines the following compactification radius for the $Q^{(i)}$\ bosons:
\begin{equation}
Q^{(i)}(z)=Q^{(i)}(z)+2\pi rh\,\,\,  \label{compattificazione1}
\end{equation}%
with$\,\ \,h\in \mathbb{Z},$ and $r$ the odd compactification radius $%
r^{2}=2pm+1$.

The primary fields for this theory are given by the vertex operators:
\begin{equation}
V^{\mathbf{a},\mathbf{b}}(z)=:e^{i\mathbf{\alpha
}_{\mathbf{a}}Q_{\mathbf{b} }(z)}:\,\ \ \ \ \ \ \ \ \ \ \ \ with\ \
\,\left( \mathbf{a},\mathbf{b}%
\right) \in \mathbb{Z}^{m}{{\times} }\mathbb{Z}^{m}
\end{equation}%
where $\alpha _{\mathbf{a}}=\mathbf{a}^{T}\mathbf{R}^{-1}$ and $Q_{\mathbf{b}%
}(z)$ is the boson with winding numbers $\mathbf{b;}$ their conformal
dimensions $h_{\mathbf{a},\mathbf{b}}$ are:
\begin{equation}
h_{\mathbf{a},\mathbf{b}}=\frac{1}{2}\left( \mathbf{a}^{T}\mathbf{R}^{-1}+%
\frac{1}{2}\mathbf{b}^{T}\mathbf{R}\right) \left( \mathbf{R}^{-1}\mathbf{a}+%
\frac{1}{2}\mathbf{Rb}\right)
\end{equation}%
The extended RCFT, the multi-components $\widehat{u(1)}_{K}^{\otimes m}$
theory, can be constructed by means of the following physical conditions
\cite{Cristofano}:

1) The topological order \cite{Wn} given by $det\left( \mathbf{K}\right) $
is an odd number and coincides with the number of primary fields of the
extended algebra \cite{Cristofano}.

2) The characters of the extended theory are closed under $\Gamma _{0}(2)$.

3) The extended theory is given by the request that its conformal blocks
correspond to primary fields with conformal dimensions which differ by a
semi-integer from those of the lowest weight states.

The extended primary fields of the $\widehat{u(1)}_{K}^{\otimes m}$ theory
are obtained by imposing that the weights are in the lattice $\mathbb{Z}_{%
\mathbf{K}}=\frac{Z^{m}}{\mathbf{K}Z^{m}}$ and the corresponding vertex
operators are:
\begin{equation}
V_{\mathbf{\lambda }}(z)=:e^{i\mathbf{\alpha }_{\lambda }{{\cdot} }\mathbf{Q}%
(z)}:
\end{equation}%
The corresponding characters are given in the standard form by:
\begin{equation}
\widetilde{\chi }_{\mathbf{\lambda }}^{\mathbf{MC}}(w|\tau )=Tr_{H_{\lambda
}}\left( q^{\left( L_{0}-\frac{m}{24}\right) }e^{2\pi iwJ}\right)
\label{car-frohli}
\end{equation}%
where $L_{0}=\frac{1}{2}\mathbf{n}^{T}\left( \mathbf{K}\right) ^{-1}\mathbf{n%
}$\ \cite{Wn} is the eigenvalue of the $L_{0}$ mode of the Virasoro algebra
and \ $J=\mathbf{t}^{T}\left( \mathbf{K}\right) ^{-1}\mathbf{n}$ is the
charge relative to $\mathbf{n}$, the vector defining the weight of the
descendent state in the Hilbert space $H_{\mathbf{\lambda }}$ of the primary
field with highest weight $\mathbf{\lambda }$. Explicitly eq.(\ref%
{car-frohli}) becomes:
\begin{equation}
\widetilde{\chi }_{\mathbf{\lambda }}^{\mathbf{MC}}(w|\tau )=\frac{e^{-\pi
m(2mp+1)\frac{\left( Imw\right) ^{2}}{Im\tau }}}{\eta \left( \tau \right)
^{m}}\sum\limits_{\mathbf{a}\in Z^{m}}e^{2\pi i\left\{ \frac{\tau }{2}\left[
\left( \mathbf{K}\mathbf{a}+\mathbf{\lambda }\right) ^{T}\left( \mathbf{K}%
\right) ^{-1}\left( \mathbf{K}\mathbf{a}+\mathbf{\lambda }\right) \right] +w%
\mathbf{t}^{T}[\left( \mathbf{K}\mathbf{a}+\mathbf{\lambda }\right)
]\right\} }  \label{u(1)k}
\end{equation}

where the non-analytic phase was introduced in \cite{Cappelli} and it is
necessary to define properly the transport of charges. The extended theory
$%
\widehat{u(1)}_{K}^{\otimes m}$ has $2pm+1$ independent primary states as it
is requested by the topological order and the characters are closed under $%
\Gamma _{0}(2)$.

\section{The symmetries of the ECFTs\label{symmetry}}

In this section we briefly analyze the symmetry properties of the different
ECFTs proposed to describe the Jain fillings.

\subsection{The $\widehat{u(1)}{\times} \widehat{su(m)}_{1}$ models}\label{sim}

The factorization of the multi-components $\widehat{u(1)}_{K}^{\otimes m}$
model in a charged and a neutral sector allows us to evidence an $\widehat{
u(1)}_{m\left( 2pm+1\right) }{{\times} }\widehat{su(m)_{1}}$ algebraic structure.
The charged sector has an $\widehat{u(1)}_{m\left( 2pm+1\right) }$ extended
symmetry\footnote{ It is defined in Appendix B. It is an odd RCFT, that is
a RCFT with respect to $\Gamma _{0}(2)$.} while the neutral one has a full
$\widehat{su(m)}_{1}$ symmetry. Moreover the two sectors are not
independent but are coupled by a discrete group so implying that a
superselection rule, called $m$-ality rule, must be taken into account in
order to build the physical excitations of the model (i.e. a holon-spinon
topological coupling).

The factorization into a charged and neutral sector is explicitly obtained
by diagonalizing the $\mathbf{K}$ matrix. The orthogonal matrix $O_{m}$,
which does it, has been explicitly given in \cite{Musto}, and the new
fields $\mathbf{Q}^{\prime }(z)= $ $O_{m}\mathbf{Q}(z)$ can be identified
with those of ref.\cite{cgm1}. We define the charged boson $X(z)\equiv
\frac{\sum_{l=1}^{m}Q_{l}^{\prime }(z)}{%
\sqrt{m}}$ and the neutral vector $\Phi(z)$ with components $Q_{i}^{\prime }(z)$ for $%
i=1,...,m-1$. The orthogonal matrix $O_{m}$ can be written in terms of the
base vectors $\mathbf{u}_{i}$ as follows:
\begin{equation}
O_{m}=\left(
\begin{array}{c}
\mathbf{u}_{1}\ ..\,\mathbf{u}_{m} \\
\frac{1}{\sqrt{m}}.\,\,.\frac{1}{\sqrt{m}}\
\end{array}%
\right) _{m{{\times} }m}
\end{equation}%
where the $m$-vectors $\mathbf{u}_{i}$ satisfy the constraint $\sum_{i}$ $%
\mathbf{u}_{i}=0$ and are related to the simple roots $\mathbf{\alpha }_{j}$
and to the fundamental weights $\mathbf{\Lambda }_{j}$ of the finite algebra
$su(m)$ by the following relations:
\begin{eqnarray}
\mathbf{\alpha }_{j} &=&\mathbf{u}_{j}-\mathbf{u}_{j+1}\,~~~;~~~~~\ \ \ \ \
\ \ \ ~\mathbf{\Lambda }_{l}=\sum_{j=1}^{l}\mathbf{u}_{j}\, \\
\mathbf{\Lambda }_{l}{{\cdot} }\mathbf{\alpha }_{j} &=&\delta
_{l}^{j}~~~~~~~;~~~~\ \ \ \ \ \ \ \ ~\mathbf{\Lambda }_{i}{{\cdot} }\mathbf{%
\Lambda }_{l}=i\left( 1-\frac{l}{m}\right) ~~\ \ ~~~~for~~\ ~~~i\leq l,\text{
\ }l=1,..,m-1.  \notag
\end{eqnarray}

We put $\phi ^{j}(z)=\mathbf{u}_{j}{{\cdot} }\mathbf{\Phi}(z) $ for $%
j=1,..,m-1$ so that the primary fields in the $\widehat{u(1)}_{K}^{\otimes
m} $ theory factorize into a charged and a neutral part:
\begin{equation}
V_{\mathbf{X }}(z)=e^{i\mathbf{\alpha }_{X}X(z)}\psi (z)\,
\end{equation}%
$\,$where $\mathbf{\alpha }_{X}=\frac{\sum_{l=1}^{m}\lambda _{l}}{\sqrt{%
m(2pm+1)}},$ $\mathbf{\lambda}_l \in \mathbb{Z}_{\mathbf{K}}$ \ and $\psi
(z)=e^{i\sum_{j=1}^{m}\lambda _{j}\phi ^{j}(z)}$.

The field $X(z)$ satisfies the following compactification condition, as a
consequence of the compactification conditions for the $\mathbf{Q}(z)$
fields in eq. $\left( \ref{compattificazione1}\right) :$
\begin{equation}
X(z)=X(z)+2\pi \,r_{X}\text{ }l\,\,\,\,\,\ \ \ where\text{ \ \ \ \ \ }r_{X}=%
\sqrt{m(2pm+1)}\,
\end{equation}

Thus the charged component of $\widehat{u(1)}_{K}^{\otimes m}$ is just the
$\widehat{u(1)}_{m\left( 2pm+1\right) }$ extended theory.

By using the neutral fields one gets the following currents:
\begin{eqnarray}
J_{\mathbf{\alpha }_{i,j}}(z) &=&c_{i,j}:e^{i(\phi ^{i}(z)-\phi
^{j}(z))}:\,\,for\,\,\mathbf{\alpha }_{i,j}=\mathbf{u}_{i}-\mathbf{u}_{j} \\
J_{h}(z) &=&i\partial \Phi _{h}(z)
\end{eqnarray}%
satisfying the $\widehat{su(m)}_{1}$ algebra\footnote{$c_{i,j}$ are cocycle
factors which allow to get the correct commutation relations.}. Let us
recall that starting from $\mathbf{\Lambda }_{j}$, $\ j=1,..,m-1$, and the
basic weight $\widehat{\mathbf{\Lambda }}_{0}$ of the affine algebra
$\widehat{su(m)}_{1}$, one can obtain all the other fundamental weights
$\widehat{\mathbf{\Lambda }}_{j}=\mathbf{\Lambda
}_{j}+\widehat{\mathbf{\Lambda }}_{0}$, $\ j=1,..,m-1$. For a general
weight $\widehat{\mathbf{\Lambda
}}=\sum_{i=0}^{m-1}n_{i}\widehat{\mathbf{\Lambda }}_{i}$ one can define the
$m$-ality charge as $l=\sum_{i=1}^{m-1}in_{i},$ $ mod(m)$. It is a good
quantum number useful to classify the excitations. As pointed out above,
the theory $\widehat{u(1)}_{K}^{\otimes m}$ is not a simple tensor product
of the charged and the neutral component. This is well evidenced by the
explicit expression of the $\widehat{u(1)}_{K}^{\otimes m}$ characters
given in eq.$\left( \ref{u(1)k}\right) ,$ which were introduced in
\cite{Musto} for the m=3 case and in \cite{Cappelli} for generic $m$ as:
\begin{equation}
\tilde{\chi}_{b}^{\mathbf{MC}}(w|\tau )=\sum_{l=0}^{m-1}\chi _{l}^{\widehat{%
su(m)}_{1}}(\tau )\bar{K}_{(2pm+1)l+mb}^{\left[ m\left( 2pm+1\right)
\right] }(w|\tau )\,  \label{carattere nontwistato carico-neutro 1}
\end{equation}%
with $b=0,..,2pm$ and where $\chi _{l}^{\widehat{su(m)}_{1}}(\tau )$ are
the characters of the $\widehat{su(m)}_{1}$ representations with integrable
highest weight (h.w.) $\widehat{\mathbf{\Lambda }}_{l}\in P_{+}^{1}$ ($%
P_{+}^{1}$ being the set of all the dominant weights of
$\widehat{su(m)}_{1}$) and conformal weight $h_l=\frac{l(m-l)}{2m}$ (i.e. a
spinon with $m$-ality charge equal to $l$). Above $\bar{K}_{c}^{\left[
m\left( 2pm+1\right)
\right] }(w|\tau )$ are the characters of the extended $%
\widehat{u(1)}_{m\left( 2mp+1\right) }$ theory and are explicitly given in
Appendix B\footnote{ Notice that the parametrization given above is not
unique due to the symmetry of the characters. They satisfy
$\tilde{\chi}_{\lambda+(2mp+1)}^{\mathbf{MC}}=\tilde{\chi}_{\lambda}^{\mathbf{MC}}$
thus there are only $2mp+1 $ independent primary fields. In \cite{Cappelli}
a parametrization which evidences the independence of the neutral component
from the flux $p$ was chosen. In this paper we give the same
parametrization for the P-P and P-A sector. Nevertheless a different
parametrization which is more useful for our approach will be given in the
A-P sector. Obviously they are completely equivalent and they can be mapped
one into the other.}.

Furthermore the form of the $\mathbf{K}$ matrix allows to consider various
discrete symmetries. Here we focus our attention on its symmetries under
the permutations of the Landau levels. They can be analyzed in the diagonal
basis in which the $\widehat{su(m)}_{1}$ symmetry is evidenced. We
distinguish two kinds of symmetries, one acting as an outer automorphism of
the $\widehat{su(m)}_{1}$ lattice and the other one associated to inner
automorphisms of the algebra (i.e. the Weyl group). Inner automorphisms
preserve the commutation relations of the algebra (although the action on
the algebra generators is nontrivial). We let $D(g)$ and $D(\hat{g})$ stand
for the symmetry group of the $su(m)$ and $\widehat{su(m)}$ Dinkin diagrams
respectively. As it is well known the group of the outer automorphisms of
$\hat{g},$ $O(\hat{g})=D(\hat{g})/D(g),$ is the dual of the center of the
group of $g$, i.e. $B(G)$ \cite{m-ality} which in the case of $su(m$) is
$Z_{m}$. Moreover it acts on the finite part of ${%
\hat{\Lambda}}$ like an element $w_{g}$ of the finite Weyl group (i.e. $g({%
\hat{\Lambda})}=(g-1){\hat{\Lambda}}_{0}+w_{g}{\hat{\Lambda}}$) by changing
the $m$-ality of a fundamental representation: $g(l)=l+1$. The $B(G)$ group
action on the algebra generators is trivial. We stress that outer
automorphisms also must preserve the commutation relations of the algebra,
so that the characters will be still given in terms of the
$\widehat{su(m)}_{1}$ ones. The extended $\widehat{u(1)}%
_{K}^{\otimes m}$ theory satisfies the important property that under the
action of $O(\hat{g})$ all its extended characters are fixed points, that
is they satisfy a gluing condition between the charged and the neutral
sector which couples a $Z_{m}$ subgroup of the $\widehat{u(1)}$ to the
$m$-ality charge in an invariant way.  Thus the P-P, periodic-periodic
sector of the TM will be invariant with respect to this coupled group (i.e.
the permutation group acts as an inner automorphism for the full theory).
This behavior is different from that of the paired states \cite{Paired2},
where the Halperin theory does not coincide with the P-P untwisted sector
of its $ Z_{m}$-orbifold. The $m$-reduction procedure induces naturally
this symmetry and gives the correct coupling between the charged and the
neutral sector \cite{cgm1}.

Furthermore quotienting by a discrete group that is not in the center it
breaks all the affine representations; therefore the partition function of
the \ TM in the twisted sector cannot be described in terms of the affine
characters (i.e. the affine symmetry is broken in these sectors to the
$W_{m} $ one).

\subsection{The $W_{m}$ Minimal model}

The ECFT description of a QHF at the plateau is justified by the
incompressibility of the Laughlin fluid. The dynamical symmetry of this
fluid in the disk geometry is the area-preserving diffeomorphisms of the
plane which imply the $W_{1+\infty }$ algebra. This is the unique centrally
extended quantum analogue of the classical area preserving diffeomorphisms
algebra $w_{\infty }$ on the circle.

The infinite generators $W_{m}^{n+1}$ of $W_{1+\infty }$ of conformal spin ($%
n+1$) are characterized by a mode index $m\in Z$ and satisfy the algebra:
\begin{equation}
\left[ W_{m}^{n+1},W_{m^{\prime }}^{n^{\prime }+1}\right] =(n^{\prime
}m-nm^{\prime })W_{m+m^{\prime }}^{n+n^{\prime }}+q(n,n^{\prime
},m,m^{\prime })W_{m+m^{\prime }}^{n+n^{\prime }-2}...+d(n,m)c\,\delta
^{n,n^{\prime }}\delta _{m+m^{\prime }=0}
\end{equation}
where the structure constants $q$ and $d$ are polynomials of their
arguments, $c$ is the central charge, and dots denote a finite number of
similar terms involving the operators $W_{m+m^{\prime }}^{n+n^{\prime }-2l}$
\cite{BS, ctz}.

Such an algebra contains an Abelian $\widehat{u(1)}$ current for $n=0$ and a
Virasoro algebra for $n=1$ with central charge $c$. Their zero modes
eigenvalues are identified as the charge and angular-momentum of the edge
excitations of the QHF.

In the literature \cite{BS,kac} the unitary representations of $W_{1+\infty
}$ have integer central charge and can be of two types: generic or
degenerate.

The generic representations, with central charge $c=m$, are equivalent to
those of the multi-components bosons $\widehat{u(1)}_{K}^{\otimes m}$, where
the $\mathbf{K}$ matrices define the lattice $\mathbb{Z}_{\mathbf{K}}=\frac{%
\mathbb{Z}^{m}}{\mathbf{K}\mathbb{Z}^{m}}$ whose weights $\mathbf{\lambda }%
=(\lambda _{1},..,\lambda _{m})$ satisfy the conditions $\lambda
_{a}-\lambda _{b}\notin \mathbb{Z}$, $\forall a\neq b\in 1,..,m$.

The degenerate representations, with central charge $c=m$, are contained
into those of the multi-components and satisfy the conditions $\lambda
_{a}-\lambda _{b}\in \mathbb{Z}$.

In particular the full degenerate representations are those whose weights
all define degenerate representations. These conditions select uniquely the
lattices $\mathbb{Z}_{\mathbf{K}}$ as the ones generated by the
$\mathbf{K}$ matrices of the Jain series. The theory with full degenerate
representation only is also called $W_{1+\infty }$ minimal model
\cite{ctz5}.

The $W_{1+\infty }$ models are isomorphic to $\widehat{u(1)}{{\times} }%
\mathcal{W}_{m}$ theory, where $\mathcal{W}_{m}$ is the algebra with central
charge $c=(m-1)$, defined by the $q\rightarrow \infty $ limit of the
Zamolodchikov-Fateev-Lukyanov algebra with $c=(m-1)\left( 1-\frac{n(n+1)}{%
q(q+1)}\right) $ \cite{fateev}. In this limit an infinity of degenerate
representations appears which are in one-to-one relation to the degenerate
representations of the above minimal models.

The $\mathcal{W}_{m}$ algebra can be also defined by a coset construction of
the kind $W(\hat{g}/g;k)$ based on the Casimir operators of a finite algebra
$g$ (see \cite{BS} and \cite{KT} for details). The relevant coset for Jain
series is $W(\widehat{su(m)}_{k}/su(m);k=1)$, which involves the finite
algebra $su(m)$, thus the central charge of the $\mathcal{W}_{m}$ has the
same value of the full theory $\widehat{su(m)}_{1}$, i.e. $c_{\mathcal{W}%
_{m}}=c_{\widehat{su(m)}_{1}}=m-1$. The h.w. representations of $\mathcal{W}%
_{m}$ are defined by decomposing those of $\widehat{su(m)}_{1}$ in terms of
 $su(m)$ ones. The characters of the coset $\mathcal{W}_{m}$ are thus
the branching functions constructed by decomposing the characters of the
affine algebra in terms of those of the finite one:
\begin{equation}
\chi _{\widehat{\mathbf{\Lambda }}_{l}}^{\widehat{su(m)}_{1}}(\xi |\tau
)=\sum_{\mathbf{\Lambda }:\widehat{\mathbf{\Lambda }}\in P_{+}^{1}\cap
\Omega _{\widehat{\mathbf{\Lambda }}_{l}}}b_{\widehat{\mathbf{\Lambda }}%
_{l}}^{\Lambda }(\tau )\chi _{\mathbf{\Lambda }}^{su(m)}(\xi )
\end{equation}%
where $\Omega _{\widehat{\mathbf{\Lambda }}_{l}}$ is the set of the weights
in the h.w. representation of $\widehat{su(m)}_{1}$\ relative to $%
\widehat{\mathbf{\Lambda }}_{l}$, the integrable h.w., $\mathbf{%
\Lambda }$ is the finite part of the affine weight $\widehat{\mathbf{\Lambda
}}$.

The above $su(m)$ characters are explicitly given by:
\begin{equation}
\chi _{\mathbf{\Lambda }}^{su(m)}(\xi )=\frac{\sum_{w\epsilon W}\epsilon
(w)e^{w(\mathbf{\Lambda }+\mathbf{\rho }){{\cdot} }\xi }}{\sum_{w\epsilon
W}\epsilon (w)e^{w(\mathbf{\rho }){{\cdot} }\xi }}
\end{equation}%
where $\mathbf{\Lambda }=\mathbf{\Lambda }_{l}+\mathbf{\gamma }$ and $\gamma
\in Q$, the set of roots of $su(m)$, $w$ is an element in the Weyl group $W$
of the $su(m)$ and $\epsilon (w)$ is its parity. Thus the characters of $%
\mathcal{W}_{m}$ are defined by $\chi _{\mathbf{\Lambda }}^{\mathcal{W}%
_{m}}(\tau )\equiv b_{\widehat{\mathbf{\Lambda }}_{l}}^{\Lambda }(\tau )$
and their explicit expressions are given by:
\begin{equation}
\chi _{\mathbf{\Lambda }}^{\mathcal{W}_{m}}(\tau )=\frac{q^{\frac{\Lambda
^{2}}{2}}}{\eta (q)^{m-1}}\prod_{\alpha \epsilon \Delta _{+}}(1-q^{(\Lambda
+\rho ){{\cdot} }\alpha })
\end{equation}

For $z\rightarrow 0$ the characters $\chi _{\mathbf{\Lambda }}^{su(m)}(%
\mathbf{\xi }=z\mathbf{\rho })$ go to the dimension $d_{su(m)}(\mathbf{%
\Lambda })$ of the $\mathbf{\Lambda }$ representation of the finite algebra $%
su(m)$. Thus the $\widehat{su(m)}_{1}$ characters can be written in the
form:
\begin{equation}
\chi _{l}^{\widehat{su(m)}_{1}}(\tau )\equiv \underset{z\rightarrow 0}{\lim }%
\chi _{\widehat{\mathbf{\Lambda }}_{l}}^{\widehat{su(m)}_{1}}(\xi =z\mathbf{%
\rho }|\tau )=\sum_{\Lambda :\widehat{\Lambda }\in P_{+}^{1}\cap \Omega _{%
\widehat{\mathbf{\Lambda }}_{l}}}d_{su(m)}(\mathbf{\Lambda })\chi _{\mathbf{%
\Lambda }}^{\mathcal{W}_{m}}(\tau )  \label{su(m)-Wm}
\end{equation}

The equivalence $\widehat{su(m)}_{1}=$ $su(m){{\times} }\mathcal{W}_{m}$
gives $\widehat{su(m)}_{1}$ as a RCFT extension of the algebra $\mathcal{W}%
_{m}$. Eq.$\left( \ref{su(m)-Wm}\right) $ shows that any integrable h.w. $%
\widehat{\mathbf{\Lambda }}_{l}$ representation of \ $\widehat{su(m)}_{1}$
is given by collecting infinite representations $\mathbf{\Lambda }$ of $%
\mathcal{W}_{m}$, such that $\widehat{\Lambda }\in P_{+}^{1}\cap \Omega _{%
\widehat{\mathbf{\Lambda }}_{l}}$, each one with multiplicity $d_{su(m)}(%
\mathbf{\Lambda })$.

Nevertheless this extension is not a minimal one which should be given by
collecting infinite primary fields of $\mathcal{W}_{m}$, everyone with
multiplicity one.

The main point for this minimal $\mathcal{W}_{m}$ theory is that it is not
a RCFT, that is the characters relative to the extended primary fields are
not closed under the modular transformation $S$. We propose a theory in
which both symmetries can be taken into account, being realized in
different sectors of the full TM. It is based on another interesting
realization of the $\mathcal{W}_{m}$ algebra given in \cite{FKN} by using
twisted bosons. The representation theory for this realization was studied
in \cite {bakalov}. This realization is the relevant one for the
$m$-reduction and introduces new topological sectors to the description of
the Jain fillings. In order to obtain the twisted characters of the
$\mathcal{W}_{m}$ algebra it is necessary to add the special vector
$\frac{\mathbf{\rho }}{m}=\frac{1}{ m}\sum_{i=1}^{m-1}\mathbf{\Lambda
}_{i},$ with $\mathbf{\rho }^{2}=$ $\frac{ m(m^{2}-1)}{12},$ to all the
representations. Nevertheless these are not representations of the algebra
$\widehat{su(m)}_1$, so this symmetry is explicitly broken in this vacuum.
This vector changes the conformal dimensions of the vacuum state according
to:
\begin{equation}
h_{\mathbf{\Lambda }}=\frac{1}{2}(\mathbf{\Lambda }+\frac{\mathbf{\rho }}{m}%
)^{2}=\frac{\mathbf{\Lambda }^{2}}{2}+\frac{m^{2}-1}{24m}+\frac{\tilde{f}}{2m%
}
\end{equation}%
so that all the $n_{i}=\delta _{i}^{l}$ are simply translated by $1/m$ with
respect to those of the minimal models. Indeed, as it follows from the
modular invariance of the theory, we find that the neutral characters for
the twisted sector can be given in terms of $\chi _{l}^{\widehat{su(m)}_{1}}(%
\mathbf{\rho }\tau /m|\tau ).$

Moreover not only the $\widehat{u(1)}{\times} \widehat{su(m)}_{1}$ is contained in
our RCFT and it coincides with the P-P sector of the TM but also an
extended $\mathcal{W}_{m}$ minimal model is realized in the P-A sector, as
it will be shown explicitly in the following sections.

\section{The m-reduction on the plane\label{planereduction}}

In this section we briefly review the $m$-reduction procedure on the plane
as a starting point for its generalization to the torus which is the main
content of this paper. Our approach is meant to describe all the plateaux
with odd denominator starting from the Laughlin filling $\nu =1/(2pm+1)$,
which is described by a CFT with $c=1$, in terms of a scalar chiral field
compactified on a circle with radius $R^{2}=1/\nu =2pm+1$ (or its dual $%
R^{2}=4/(2pm+1)$). Then the $u(1)$ current is given by $J(z)=i\partial
_{z}Q(z)$, where $Q(z)$ is the compactified Fubini field with the standard
mode expansion:
\begin{equation}
Q(z)=q-i\,p\,lnz+\sum_{n\neq 0}\frac{a_{n}}{n}z^{-n}  \label{modes}
\end{equation}%
with $a_{n}$, $q$ and $p$ satisfying the commutation relations $\left[
a_{n},a_{n^{\prime }}\right] =n\delta _{n,n^{\prime }}$ and $\left[ q,p%
\right] =i$.

The primary fields are expressed in terms of the vertex operators $U^{\alpha
_{s}}(z)=:e^{i\alpha _{s}Q(z)}:$ with $s=1,...,2pm+1$ and conformal
dimension $h=\frac{s^{2}}{2(2pm+1)}$. The $h=pm+\frac{1}{2}$ field
($s+2pm+1$) describes the electron with electric charge $q_{e}=1$ in units
of the electron charge $e$ and magnetic one $q_{m}=2pm+1$ in unit of
$\frac{hc}{e}$, while to other $2pm$ primary fields correspond to anyons
with lowest charges $q_{e}=\frac{1}{2pm+1}$ and $q_{m}=1.$ The dynamical
symmetry is given by the $W_{1+\infty }$ algebra with $c=1$, whose
generators are simply given by a power of the current $J(z)$.

We start with the set of fields in the above CFT (mother theory). Using the $%
m$-reduction procedure, which consists in considering the subalgebra
generated only by the modes in eq.(\ref{modes}), which are multiple of an
integer $m$, we get the image of the twisted sector of a $c=m$ orbifold CFT
(daughter theory, i.e. the TM) which describes the Lowest Landau Level (LLL)
dynamics of the new filling $\nu =m/(2pm+1)$ .

Then the fields in the mother CFT can be factorized into irreducible orbits
of the discrete $Z_{m}$ group which is a symmetry of the TM and can be
organized into components which have well defined transformation properties
under this group. To compare the orbifold so built with the $c=m$ CFT, we
use the mapping $z\rightarrow z^{1/m}$and the isomorphism defined in ref.
\cite{VM} between fields on the $z$ plane and fields on the $z^{m}$ covering
plane given by the following identifications: $a_{nm+l}\longrightarrow \sqrt{%
m}a_{n+l/m}$, $q\longrightarrow \frac{1}{\sqrt{m}}q$.

We perform a \textquotedblleft double\textquotedblright -reduction which
consists in applying this technique into two steps.

1) The $m$-reduction is applied to the Fubini field $Q(z).$ This induces
twisted boundary conditions on the currents. It is useful to define the
invariant scalar field:
\begin{equation}
X(z)=\frac{1}{m}\sum_{j=1}^{m}Q(\varepsilon ^{j}z)  \label{X}
\end{equation}%
where $\varepsilon ^{j}=e^{i\frac{2\pi j}{m}}$, corresponding to a
compactified boson on a circle with radius now equal to $%
R_{X}^{2}=R^{2}/m=2p+1/m$. This field describes the $u(1)$ electrically
charged component of the new filling.

On the other hand the non-invariant fields defined by
\begin{equation}
\phi ^{j}(z)=Q(\varepsilon ^{j}z)-X(z),~~~~~~~~~~~~~~~~\sum_{j=1}^{m}\phi
^{j}(z)=0
\end{equation}%
naturally satisfy twisted boundary conditions so that the $J(z)$ current of
the mother theory decomposes into a charged current given by $J(z)=i\partial
_{z}X(z)$ and $m-1$ neutral ones $\partial _{z}\phi ^{j}(z)$ \cite{cgm1}.

2) The $m$-reduction applied on the vertex operators $U^{\alpha _{s}}(z)$ of
the mother theory also induces twisted boundary conditions on the vertex
operators of the daughter CFT. The discrete group used in this case is just
the $m$-ality group.

The vertex operator in the mother theory can be factorized into a vertex
that depends only on the invariant field:
\begin{equation}
\mathcal{U}^{\alpha _{s}}(z)=z^{\frac{\alpha _{s}^{2}(m-1)}{m}}:e^{i\alpha
_{s}{\ }X(z)}:
\end{equation}%
and in vertex operators depending on the $\phi ^{j}(z)$ fields. We
introduce the neutral component:
\begin{equation}
\psi _{1}(z)=\frac{z^{\frac{1-m}{m}}}{m}\sum_{j=1}^{m}\varepsilon
^{j}:e^{i\phi ^{j}(z)}:
\end{equation}%
which invariant under the twist group given in 1) and has $m$-ality charge
$l=1$.

The set of primary fields generated by their product can be given in terms
of the fundamental representations $\Lambda _{l}$ of the $su(m)$ Lie
algebra. By looking at their conformal dimension we are led to identify
them as $%
\widehat{su(m)}_{1}$ spinons given in a $m$-ality diagonal basis. Indeed we
can define the full set of fields:
\begin{equation}
\psi _{l}(z)=\sum_{j_{1}>j_{2}>\dots >j_{l}}f(\varepsilon ^{j_{1}},\dots
,\varepsilon ^{j_{l}},z):e^{i\phi ^{j_{1}}(z)}\dots e^{i\phi ^{j_{l}}(z)}:
\label{spinons}
\end{equation}%
where the functions $f(\varepsilon ^{j_{1}},\dots ,\varepsilon ^{j_{l}},z)$
can be extracted from the OPE relations. The sum takes into account the fact
that any field can be associated to the $l$-th fundamental representation of
$su(m)$ (namely, the antisymmetric tensor representation) and then have $m$%
-ality $l$.

Higher spin currents in $W_{1+\infty }$ algebra are given by the infinite
generators in the enveloping algebra of the $u(1)$ charged sector and by the
$\mathcal{W}_{m}$ currents obtained from the neutral sector. The explicit
form for $m\leq 4$ was given in \cite{FKN}. We report here only the first
element beyond the spin two of the series which in our basis is expressed
as:
\begin{equation}
W^{3}(z)=\frac{1}{2\sqrt{m}}\sum_{j,j^{\prime },j^{\prime \prime
}=1}^{m}:\partial _{z}\phi ^{j}(z)\partial _{z}\phi ^{j^{\prime
}}(z)\partial _{z}\phi ^{j^{\prime \prime }}(z):
\end{equation}

All the fields given in eq.(\ref{spinons}) are invariant under the $Z_{m}$
group twisting the $\phi ^{j}(z)$ fields. The remaining primary fields in
the neutral sector which are not invariant are given by:
\begin{equation}
\bar{\psi}_{g}(z)=\frac{z^{\frac{(1-m)}{m}}}{m}\sum_{j=1}^{m}\varepsilon
^{gj}:e^{i\phi ^{j}(z)}:
\end{equation}%
with $g=2,...,m$. This decomposition breaks explicitly the finite $su(m)$
symmetry of the multi-components model. As we will show in the next section
the residual symmetry of the twist invariant sector is just the $W_{1+\infty
}$ one.

\bigskip

\section{\label{torusreduction}The m-reduction on the torus}

In the following we consider the case of prime $m$ only, the more general
case will be the subject of a forthcoming paper.

\subsection{The TM for prime $m>2$\label{m-rid}}

The present discussion is valid for any prime $m>2$ while in sec.(\ref{TM=2}
) it is analyzed the special $m=2$ case. This is not only to clarify the
construction in the simplest case but also for the peculiarity of the $m=2$
case in which the $m$-reduction furnishes special results.

The extended TM theory is the realization on the torus topology of that
proposed in \cite{cgm1} to describe the plateaux for the Jain series $\nu
=%
\frac{m}{2pm+1}$ on the plane. We can obtain the full theory by applying the
$m$-reduction on the torus and by implementing the modular invariance
condition. The mother theory describes the Laughlin filling $\nu
=\frac{1}{ 2mp+1}$ corresponding to the $\widehat{u(1)}_{2mp+1}$ symmetry
(see appendix B for details).

By applying the $m$-reduction on the fields in the mother theory we obtain
the twisted sector of the TM immediately. In this procedure it is necessary
to take into account two points:

1) The $\widehat{u(1)}_{2pm+1}$ mother theory is not an ordinary RCFT but a
theory which is invariant only under $\Gamma _{0}(2)$.

2) The TM is a fermionic theory so it must be of the same type of the
mother theory (i.e. the conformal dimensions of its primary fields are
defined only $mod(1/2)$).

The extended primary fields in the twisted sector have conformal dimensions%
\footnote{%
We remember that the parametrization here used for the untwisted characters
is different from that of the twisted one. }:
\begin{equation}
h_{(s,\tilde{f})}=\frac{s^{2}}{2(2pm+1)m}+\frac{m^{2}-1}{24m}+\frac{\tilde{f}%
}{2m}\,
\end{equation}%
and are parametrized by $\ s=0,..,2pm$, which is the quantum number
distinguishing the fields in the mother theory and by $\tilde{f}=0,..,m-1$,
which indicates the twisted sector in the model. The term $\frac{m^{2}-1}{24m%
}$ is due to the twisted conditions. The condition 2) implies that the
number of conformal blocks of the extended theory are $m$ instead of $2m$.
The corresponding characters can be easily given in terms of those of the
mother theory:
\begin{equation}
\chi _{(s,\tilde{f})}(w|\tau )=\frac{1}{m}\sum_{j=0}^{m-1}e^{-\frac{2\pi i}{m%
}\left( 2j\right) (\frac{\tilde{f}}{2}+\frac{s^{2}}{2(2pm+1)}-\frac{1}{24})}%
\bar{K}_{s}^{\left( 2pm+1\right) }(w|\frac{\tau +2j}{m})
\label{carattere twistato}
\end{equation}

The closure under $\Gamma _{0}(2)$ of the TM allows us to obtain part of the
fields in the untwisted sector. We get $2pm+1$ \ new primary fields for this
sector with conformal dimensions:

\begin{equation}
\tilde{h}_{b}=\frac{(mb)^{2}}{2(2pm+1)m}\,
\end{equation}%
They can be identified with the $m$-particles introduced in
\cite{Cristofano} and are described by the characters
\begin{equation}
\tilde{\chi}_{b}(w|\tau )=\frac{1}{\sqrt{m}}\bar{K}_{b}^{\left( 2pm+1\right)
}(mw|m\tau )  \label{carattere nontwistato}
\end{equation}%
Their peculiarity is that the neutral part of the vertex operator is the
identity. It gives a contribution only by means of its descendents which
are just the $\mathcal{W}_m $ currents.

Although these fields are sufficient to satisfy the $\Gamma _{0}(2)$
covariance they are not the complete set of fields in the TM. As for the
paired states case analyzed in \cite{cgm2}, the periodic-periodic sector of
the theory can be obtained by exploring the contents of the orbifold theory $%
\widehat{u(1)}_{K}^{\otimes m}/Z_{m}$, which is the daughter CFT we obtained
by $m$-reduction. Therefore in the next section we study this orbifold in
detail and extract the information we need to complete the model.

\subsubsection{\label{Gamma _(2)}The orbifold $\widehat{u(1)}_{K}^{\otimes
m}/Z_{m}$}

Starting with the $\widehat{u(1)}_{K}^{\otimes m}$ theory, which is
invariant under the discrete symmetries analyzed in sec.(\ref{sim}), we
perform an orbifold construction with\ respect to the discrete $Z_{m}$
group. It should be noticed that the $\widehat{u(1)}_{K}^{\otimes m}$ is a $%
\Gamma _{0}(2)$-RCFT so the generated orbifold will be of the same kind.

All the primary fields of the extended $\widehat{u(1)}_{K}^{\otimes m}$
theory correspond to diagonal weights $\mathbf{b}=b\mathbf{t}$. Defining
$g$ as the generator of the $Z_{m}$ group, permuting the levels, we see
that all these fields are invariant under its action and are fixed points
with respect to this symmetry. The stabilizer $S_{b}$ for a general field
corresponding to the weight $\mathbf{b}$ is just $Z_{m}$, i.e. $%
S_{b}=\left\{ g^{i}\,:\,\ i=0,..,m-1\right\} $ and it has $m$ elements.

The untwisted P-A sector of the orbifold is obtained introducing a new
primary field for any element $\pi \in $ $S_{b}$ and any field in the mother
theory. The generic conformal field is parametrized by the index $\mathbf{b}$
and an element $\pi \in $ $S_{b}$ and its conformal dimension given by:%
\begin{equation}
\tilde{h}_{(b,\left( \mathbf{1,}\pi \right) )}=\frac{(mb)^{2}}{2(2pm+1)m}\,
\label{dim nontwist orb}
\end{equation}%
for $\pi \neq g^{0}$. The corresponding characters are%
\begin{equation}
\tilde{\chi}_{(b,\left( \mathbf{1,}\pi \right) )}(w|\tau )=\left\{
\begin{array}{c}
\tilde{\chi}_{b}^{\mathbf{MC}}(w|\tau )\text{ \ \ \ \ \ \ \ \ \ \ \ \ \ } \\
\bar{K}_{b}^{\left( 2pm+1\right) }(mw|m\tau )%
\end{array}%
\begin{array}{c}
For\text{ \ \ \ \ \ }\pi =g^{0}=\mathbf{1}\text{ \ \ \ \ \ \ \ \ \ \ \ \ \ \
\ \ \ \ \ \ \ \ } \\
For\,\text{\ \ \ \ }\pi =g^{i}\,,\text{\ \ \ }i=1,..,m-1\,\ \ \ \
\end{array}%
\right.   \label{carattere nontwistato orb}
\end{equation}%
Eq.(\ref{carattere nontwistato orb}) implies that the conformal block
relative to the identity coincides with that of the $\widehat{u(1)}%
_{K}^{\otimes m}$ theory while there is a $m-1$ degeneracy in the other
blocks. The characters $\chi _{(b,\left( \mathbf{1,}g^{i}\right) )}(w|\tau )$
of the orbifold are equivalent to the characters $\sqrt{m}\tilde{\chi}%
_{b}(w|\tau )$ given in eq.(\ref{carattere nontwistato}).

The twisted A-P sector of the $\widehat{u(1)}_{K}^{\otimes m}/Z_{m}$ model
can be obtained by the request of covariance under $\Gamma _{0}(2)$. Indeed
from $\tilde{\chi}_{(s,\left( \mathbf{1,}g^{i}\right) )}(w|\tau )$ we can
generate the characters of the twisted sector by means of the
transformation $S$. The closure under $T^{2}$ implies that also the
characters $\chi_{(s,\left( g^{i},g^{j}\right) )}(w|\tau
)=\bar{K}_{s}^{\left( 2pm+1\right) }(w|\frac{\tau +2j}{m}),$\ for
$j=1,..,m-1,$ belong to this sector\footnote{ The closure is given when
$2j\in $ $0,2(m-1)$ because the characters $\bar{K}%
_{\lambda }^{\left( 2pm+1\right) }(w|\tau )$, of $\widehat{U(1)}_{2pm+1}$,
are closed under $T^{2}$ and not under $T$.}.

The full twisted sector of this $\widehat{u(1)}_{K}^{\otimes m}/Z_{m}$ \
orbifold is given then by the characters
\begin{equation}
\chi _{(s,\left( g^{i},g^{j}\right) )}(w|\tau )=\bar{K}_{s}^{\left(
2pm+1\right) }(w|\frac{\tau +2j}{m})  \label{carattere twistato orb}
\end{equation}%
for $s=0,..,2pm$,$\,\ i=1,..,m-1{\ }${and \ }$j=0,..,m-1$. These characters
are degenerate with respect to the index $i$ as it was pointed out before.

As we have seen the P-P sector of this orbifold coincides with the $\widehat{%
u(1)}_{K}^{\otimes m}$ itself. The remaining part of the untwisted sector
coincides with the untwisted sector of the theory built by the $m$-reduction
procedure, when we take into account the $m-1$ degeneracy. The twisted A-P
sector is given by characters of eq.(\ref{carattere twistato orb}). Looking
at eq. ( \ref{carattere twistato orb}) and eq.(\ref{carattere twistato}), we
find the relationship between these characters and the characters obtained
by $m$-reduction. Indeed as a consequence of the invertibility of the
equality (\ref{carattere twistato}) for any fixed $j$, the characters in
eqs.(\ref{carattere twistato orb}) and (\ref{carattere twistato}) simply
define two different basis of the same twisted sector.

Thus we can conclude that the $m$-reduction procedure generates the non
periodic part of the untwisted sector (P-A) and the twisted sector (A-P) of
the $Z_{m}$-orbifold of $\widehat{u(1)}_{K}^{\otimes m}$, by considering
that any extended primary field has degeneracy $m-1$.

\subsubsection{Charged and neutral components of the characters of the TM}

In ref.\cite{cgm1} the TM was given in terms of charged and neutral
components on the plane. There it was shown that the charged component
coincides with the compactified theory with $R^{2}=\left( 2mp+1\right)
/m=1/\nu $. In the torus topology this separation can be given by the
decomposition of the characters in terms of those of the charged and
neutral ones. This decomposition also tells us the way in which the primary
fields of the charged and neutral components must combine in order to
generate the extended primary fields of the full TM theory.

For the P-P sector the separation is given in eq.(\ref{carattere
nontwistato carico-neutro 1}), while for the P-A sector of the TM we obtain
\begin{equation}
\tilde{\chi}_{b}(w|\tau )=\tilde{N}(\tau )\bar{K}_{mb}^{\left[ m\left(
2pm+1\right) \right] }(w|\tau )\,
\label{carattere nontwistato carico-neutro 2}
\end{equation}%
where $\tilde{N}(\tau )=\frac{1}{\sqrt{m}}\frac{\eta (\tau )}{\eta (m\tau )}$
represents the neutral contribution.

To clarify the relationship between the TM and the $\mathcal{W}_{m}$ minimal
model we point out that the twist group acts only on the $su(m)$, while the $%
\mathcal{W}_{m}$ algebra is unaffected under it. Therefore we can write ($%
\widehat{su(m)}_{1}/Z_{m}=$ $su(m)/Z_{m}{{\times} }\mathcal{W}_{m}$) and the
effect of the twist is only in the reduction of the degeneracy of the $%
\mathcal{W}_{m}$ representations as it happens in the minimal models.

Nevertheless, while the characters of $\widehat{su(m)}_{1}$ transform
covariantly under the modular group, the $\mathcal{W}_{m}$ ones do not
close under this group. That can be easily understood if one notes that the
$ SL(2,Z)$ also acts on the representations of the finite $su(m)$ algebra.
Therefore one cannot have a covariance of the $\mathcal{W}_{m}$ characters
without a covariance of the degenerate vacuum states of dimension
$d(\mathbf{\Lambda }_{a}+\mathbf{\gamma })$. To do that we relate a finite
subgroup of the modular group to the $su(m)$ automorphisms.

By using the identity
\begin{equation}
\frac{\eta (\tau )^{m}}{\eta (m\tau )}=\frac{\prod_{j=1}^{m}\Theta
_{3}\left( w-\frac{j}{m}|\tau \right) }{\Theta _{3}\left( mw-\frac{m+1}{2}%
|m\tau \right) }
\end{equation}%
we give the neutral sector in terms of the characters of $\widehat{su(m)}_{1}%
\footnote{%
Let us recall that they are not characters of $\widehat{su(m)}_{1}$ and
then we explicitly break the $su(m)$ symmetry.}$:
\begin{equation*}
\tilde{N}(\tau )=\frac{1}{\sqrt{m}}\chi _{0}^{\widehat{su(m)}_{1}}(\mathbf{%
\rho }/m|\tau )
\end{equation*}%
where $\mathbf{\rho }$ is the Weyl vector of the algebra. In this way
eq.(\ref{carattere nontwistato carico-neutro 2}) can be related to the
$\widehat{su(m)}_1$ characters of the P-P sector.

For the twisted sector the separation into charged and neutral components is
given by\footnote{%
Notice that the m-ality charge in the twisted sector is $l=a+s$ $mod$ $(m)$}:

\begin{equation}
\chi _{(s,\tilde{f})}(w|\tau )=\sum_{a=0}^{m-1}N_{(a,(s,\tilde{f}%
))}^{(p)}(\tau )\bar{K}_{(2pm+1)a+s}^{\left[ m\left( 2pm+1\right) \right]
}(w|\tau )  \label{twistato carico-neutro m}
\end{equation}%
where%
\begin{equation}
\,N_{(a,(s,\tilde{f}))}^{(p)}(\tau )=\frac{1}{m}\sum_{j=0}^{m-1}e^{-\frac{%
2\pi i}{m}\left( 2j\right) (\frac{\tilde{f}}{2}+\frac{s^{2}}{2(2pm+1)}-\frac{%
1}{24}-\frac{\left[ (2pm+1)a+s\right] ^{2}}{2(2pm+1)})}\frac{\eta (\tau )}{%
\eta (\frac{\tau +2j}{m})}
\end{equation}%
We must observe that the neutral part does not depend on $p$, that is on
the flux attached to the charged component. That can be seen by showing
that the characters of the neutral part, defined in the above formula, can
be written in terms of the ones with $p=0$ as:
$N_{(a,(s,\tilde{f}))}^{\left( p\right) }(\tau )=N_{(a,\tilde{f}^{\prime
})}^{\left( 0\right) }(\tau ),$ where $%
\tilde{f}^{\prime }=\tilde{f}-2as$ $mod$ $(m)$.

Looking at eqs.(\ref{carattere nontwistato carico-neutro
1}),(\ref{carattere nontwistato carico-neutro 2}) and (\ref{twistato
carico-neutro m}) and taking into account the parametrization change, we
conclude that the charged component is the same in any sector of the TM and
it coincides with the extended theory $\widehat{u(1)}_{m(2mp+1)}$.

Let us notice that the characters given here are written in a basis which
is explicitly invariant under the permutation of the Landau levels. In this
basis we do not see the excitations content of the theory. We could
introduce another basis in which the $m$-ality is diagonal, evidencing the
primary fields with conformal dimensions $h_{(l,b)}=$ $\frac{%
((2pm+1)l+mb)^{2}}{2m(2pm+1)}+\frac{l(m-l)}{m}$ which are the charged
excitations corresponding to the $l$-electrons ($b=0$) and the $l$-anions
($ b\neq 0$) which are fixed under the twist group. Moreover in the twisted
sector there are also neutral excitations with conformal dimensions
$\bar{h}= $ $\frac{m^{2}-1}{24m}$ (i.e. twist operators) which are a
characteristics of the TM and it was not present in the models given in
\cite{Cappelli, Frohlich}. They were independently introduced in
\cite{cgm1} and \cite{PS} giving rise to a clustering phenomenon in the
correlator of N electrons (see \cite{cgm1} for details). The role of the
twisted sector has been recently clarified for the paired states in
\cite{impurity}. In that description the neutral current was shown to
originate from a point-like interaction between the layers, which can be
attributed to the presence of a localized impurity. We will give a similar
analysis in a forthcoming paper.

\subsection{The TM for $m=2$\label{TM=2}}

For $m=2$ the TM is built following the same construction made in \ref{m-rid}%
, for the case $m>2$ and prime.

The content of the twisted sector is given by the $m$-reduction procedure
applied to the mother theory with symmetry $\widehat{u(1)}_{\left(
4p+1\right) }$. The extended primary fields have the conformal dimensions
\begin{equation}
h_{(s,\tilde{f})}=\frac{s^{2}}{4\left( 4p+1\right) }+\frac{1}{16}+\frac{%
\tilde{f}}{4}\,
\end{equation}%
$\,$with $s=0,..,4p,${\ }$\tilde{f}=0,1$, and the corresponding characters
are given by
\begin{equation}
\chi _{(s,\tilde{f})}(w|\tau )=K_{2(\left( 4p+1\right) \tilde{f}+s)}^{\left(
2\left( 4p+1\right) \right) }(w|\frac{\tau }{2})
\label{carattere twistato-m=2}
\end{equation}%
where $K_{q}^{\left( 2\left( 4p+1\right) \right) }(w|\tau )$ are the
characters of the ordinary\footnote{ See Appendix B.} RCFT
$\widehat{u(1)}_{2\left( 4p+1\right) }$.

The theory $\widehat{u(1)}_{K}^{\otimes 2}$ is given by the P-P sector of
the TM. The remaining part of its untwisted sector, that is the P-A part, is
generated by the request of closure under $\Gamma _{0}(2)$. Thus the
extended primary fields of the P-A part are obtained by performing the $S$
modular transformation on the characters of the extended primary fields of
the twisted sector. They have the following conformal dimension
\begin{equation}
\tilde{h}_{(\mu ,g)}=\frac{(\left( 4p+1\right) g+\mu )^{2}}{4(4p+1)}
\end{equation}%
where $\mu =0,..,4p${,}${\;g=}0,1$, with correspondent characters
\begin{equation}
\tilde{\chi}_{(\mu ,g)}(w|\tau )=\frac{1}{\sqrt{2}}\left( K_{\left(
4p+1\right) g+\mu }^{\left( 2\left( 4p+1\right) \right) }(2w|2\tau
)+K_{2\left( 4p+1\right) +\left( 4p+1\right) g+\mu }^{\left( 2\left(
4p+1\right) \right) }(2w|2\tau )\right)
\end{equation}

We give now their separation into charged and neutral components and relate
it to known results in the c=1 orbifold theory literature \cite{DVVV}.

The characters of the twisted sector can be written as:
\begin{equation}
\chi _{(s,\tilde{f})}(w|\tau )=\chi _{0}^{\widehat{su(2)}_{1}}(\frac{\mathbf{%
\rho }\tau }{2}|\tau )\bar{K}_{\left( 4p+1\right) \tilde{f}+s}^{(2\left(
4p+1\right) )}(w|\tau ),  \label{carico-neutro twistato m=2}
\end{equation}

while for the P-P sector eq\textbf{.} (\ref{carattere nontwistato
carico-neutro 1}) gives the following separation:%
\begin{equation}
\tilde{\chi}_{b}^{\mathbf{MC}}(w|\tau )=\chi _{0}^{\widehat{su(2)}_{1}}(\tau
)\bar{K}_{2b}^{\left( 2\left( 4p+1\right) \right) }(w|\tau )\,+\chi _{1}^{%
\widehat{su(2)}_{1}}(\tau )\bar{K}_{\left( 4p+1\right) +2b}^{\left( 2\left(
4p+1\right) \right) }(w|\tau )\,\,\,  \label{U(1)-SU(2)-m=2}
\end{equation}%
$b=0,..,4p$. For the remaining P-A sector we get
\begin{equation}
\tilde{\chi}_{(\mu ,g)}(w|\tau )=\chi _{0}^{\widehat{su(2)}_{1}}(\frac{%
\mathbf{\rho }}{2}|\tau )\bar{K}_{\left( 4p+1\right) g+\mu }^{(2\left(
4p+1\right) )}(w|\tau )\,  \label{nontwistato carico-neutro m=2}
\end{equation}%
where
\begin{equation}
\chi _{0}^{\widehat{su(2)}_{1}}(\frac{\mathbf{\rho }}{2}|\tau )=\frac{\Theta
_{4}(0|2\tau )}{\eta }
\end{equation}

Eqs.$(\ref{U(1)-SU(2)-m=2})$, $(\ref{nontwistato carico-neutro m=2})$ and $(%
\ref{carico-neutro twistato m=2})$ show that the charged component of the TM
for $m=2$ is the extended theory $\widehat{u(1)}_{2(4p+1)}$ which is an
ordinary RCFT, closed under the full modular group. It justifies the
particular modular behavior of the characters of the P-A sector which is
closed under the modular transformation $T$. It is due to the bosonic nature
of the 2-electron. It is worth noticing that even in such a case the full TM
theory remains a $\Gamma _{0}(2)$ RCFT.

Finally we notice that the P-A sector is bigger than the theory built by the
action of the generator of the discrete symmetry group $Z_{2}$. This can be
seen by writing the results of section \ref{m-rid} for $m=2$. The comparison
of eq.$\left( \ref{carattere nontwistato carico-neutro 2}\right) $,
evaluated at $m=2,$ i.e.
\begin{equation}
\tilde{\chi}_{b}(w|\tau )=\chi _{0}^{\widehat{su(2)}_{1}}(\frac{\mathbf{\rho
}}{2}|\tau )\bar{K}_{2b}^{(2\left( 4p+1\right) )}(w|\tau )\text{ \ \ \ \ \
\newline
\ \ }b=0,..,4p  \label{nontwistato c-n m=2}
\end{equation}%
with the correspondent of eq.$\left( \ref{nontwistato carico-neutro m=2}%
\right) $, makes it clear. Indeed, eq.$\left( \ref{nontwistato c-n m=2}%
\right) $ describes exactly half of the characters given in eq.$\left( \ref%
{nontwistato carico-neutro m=2}\right) $.

This anomalous behavior can be understood by noticing that the $c=1$ $Z_{2}$%
-orbifold has an extra symmetry due to a $h=1$ current (cos $\phi $). Its
radius corresponds to the multicritical point in which orbifold and
Gaussian theory coincide. The Gaussian formulation can be obtained by means
of a non-linear transformation as it was given in ref.\cite{priadko}.

The field content of the neutral sector can be easily deduced by the well
know spectrum of the $c=1$ $Z_{2}$-orbifold at the radius $R^{2}=4$. They
are: $I$\ ($h=0$), $\partial \phi $ ($h=1$), $\psi $ ($h=\frac{1}{4}$), $%
\bar{\psi}$ ($h=\frac{1}{4}$), $\sigma _{i}$ ($h=\frac{1}{16}$), $\tau _{i}$
($h=\frac{9}{16}$), $\psi $ and $\bar{\psi}$ are two kinds of
$\widehat{su(2)}_{1}$ spinons. They can be immediately related to the
standard basis as follows: $%
\psi =\psi ^{(1)}+\psi ^{(2)}$; $\bar{\psi}=\psi ^{(1)}-\psi ^{(2)}$.

Differently from the paired states case the fields $\psi $ and $\bar{\psi}$
are not independent so they do not realize two independent CFTs as it
happens in that case. They instead realize Gentile parafermions. As it is
well known, a Gentile parafermion gives a $\frac{m-1}{m}$ contribution to
the central charge. We need to have $m$ of such a parafermions to build a
unitary CFT with integer charge $c=m-1$. As a consequence their fusion rules
are: $\psi {{\times} }\psi =\partial \phi $; $\psi {{\times} }\bar{\psi}=I$;
$\bar{\psi}{{\times} }\bar{\psi}=\partial \phi $.

Due to the lack of zero modes in the twisted current the pseudospin $%
J_{3}^{su(2)}=m=(N^{(1)}-N^{(2)})/2$ is zero (or 1/2) so that the ground
state has an equal number of $\psi ^{(1)}$ and $\psi ^{(2)}$ states for $m=0$
 or $N^{(1)}=N^{(2)}+1$ for $m=1/2$. While $\psi $ increases the
($m$-ality) $l\rightarrow l+1$, $\bar{\psi}$ acts only on the third
component as $ m\rightarrow m+1/2$.

\section{The TM partition function\label{partitionfunction}}

The diagonal partition function of the part of the TM generated only by the $%
m$-reduction procedure, without taking care of the $m-1$ degeneracy, is
\begin{equation}
Z^{\left( m\right) }(w|\tau )=Z_{untwist}^{\left( m\right) }(w|\tau
)+Z_{twist}^{\left( m\right) }(w|\tau )  \label{Z-m-Ridotta}
\end{equation}%
where\footnote{%
In the $m=2$ case all is the same with the only exception that
\begin{equation}
Z_{untwist}^{\left( 2\right) }(w|\tau )=\sum_{\mu
=0}^{4p}\sum_{g=0}^{1}\left\vert \tilde{\chi}_{(\mu ,g)}(w|\tau )\right\vert
^{2}
\end{equation}%
}
\begin{equation}
Z_{untwist}^{\left( m\right) }(w|\tau )=\sum_{b=0}^{2pm}\left\vert \tilde{%
\chi}_{b}(w|\tau )\right\vert ^{2}\ \ ;\ \ \ \ \ \ Z_{twist}^{\left(
m\right) }(w|\tau )=\sum_{\tilde{f}=0}^{m-1}\sum_{s=0}^{2pm}\left\vert \chi
_{(s,\tilde{f})}(w|\tau )\right\vert ^{2}
\end{equation}%
The diagonal partition function of the TM, as a consequence of its
definition as a $Z_{m}$-orbifold, is
\begin{equation}
Z_{TM}(w|\tau )=\frac{1}{\left\vert Z_{m}\right\vert }\sum_{\forall (\pi
_{1},\pi _{2})\in Z_{m}{{\times} }Z_{m}}Z_{(\pi _{1},\pi _{2})}^{\widehat{%
u(1)}_{K}^{\otimes m}}\left( w|\tau \right) =\frac{1}{m}\sum_{i=0}^{m-1}%
\sum_{j=0}^{m-1}Z_{(g^{i},g^{j})}^{\widehat{u(1)}_{K}^{\otimes m}}\left(
w|\tau \right)
\end{equation}%
where we have defined
\begin{equation}
Z_{(g^{i},g^{j})}^{\widehat{u(1)}_{K}^{\otimes m}}\left( w|\tau \right)
=\sum_{s=0}^{2pm}\left\vert \tilde{\chi}_{(s,\left( g^{i},g^{j}\right)
)}(w|\tau )\right\vert ^{2}\,\,
\end{equation}%
$i\in \left\{ 0,..,m-1\right\} ,j\in \left\{ 0,..,m-1\right\} .$ In
particular the diagonal partition function of the P-P sector coincides with
that of the $\widehat{u(1)}_{K}^{\otimes m}$ theory%
\begin{equation}
Z_{(g^{0},g^{0})}^{\widehat{u(1)}_{K}}\left( w|\tau \right)
=\sum_{b=0}^{2pm}\left\vert \tilde{\chi}_{b}^{\mathbf{MC}}(w|\tau
)\right\vert ^{2}  \label{Z-p-p}
\end{equation}%
Furthermore following the equalities:
\begin{equation}
Z_{untwist}^{\left( m\right) }(w|\tau )\,=\frac{1}{m}Z_{(\mathbf{1},g^{i})}^{%
\widehat{u(1)}_{K}^{\otimes m}}\left( w|\tau \right) ;\,\text{\ \ \ \ \ \ \ }%
Z_{twist}^{\left( m\right) }(w|\tau )\,=\frac{1}{m}%
\sum_{h=0}^{m-1}Z_{(g^{i},g^{h})}^{\widehat{u(1)}_{K}^{\otimes m}}\left(
w|\tau \right)  \label{Z-generali}
\end{equation}%
the TM partition function can be finally rewritten as:
\begin{equation}
Z_{TM}(w|\tau )=\frac{1}{m}Z_{\widehat{u(1)}_{K}^{\otimes m}}\left( w|\tau
\right) +\left( m-1\right) Z^{\left( m\right) }(w|\tau )
\end{equation}

\section{The symmetries of the TM sectors\label{discussion}}

In this section we will discuss the relationship between the TM and the $%
\widehat{u(1)}_{m\left( 2pm+1\right) }{{\times} }\mathcal{W}_{m}$ minimal
models in detail.

As we have already seen the minimal model is not a RCFT. The fact that the
TM is the $Z_{m}$-orbifold of the $\widehat{u(1)}_{K}^{\otimes m}$ model
clarifies the relation between the two previous proposed theories for the
Jain plateaux: the multi-component bosonic theory, characterized by the
extended $\widehat{u(1)}_{m\left( 2pm+1\right) }{{{\times} } }\widehat{%
su(m)_{1}}$ algebra and the extended minimal model.

It is very easy to see that the $\widehat{u(1)}_{m\left( 2pm+1\right) }{{%
{\times} }}\widehat{su(m)_{1}}$ algebra is the symmetry of the P-P sector of the
TM. It is a non-minimal sector with respect to the $\mathcal{W}_{m}$
symmetry. To see that we can write, in the particular $m=2$\ case, eq.$%
\left( \ref{su(m)-Wm}\right) $ which gives the characters of the affine $%
\widehat{su(2)}_{1}$ in terms of the $\mathcal{W}_{2}$\ characters:
\begin{equation}
\chi _{l}^{\widehat{su(2)}_{1}}(\tau )=\sum_{n=0}^{\infty }d_{su(2)}(n+\frac{%
l}{2})\chi _{2n+l}^{\mathcal{W}_{2}}(\tau )
\end{equation}%
where $l= 0,1$. They correspond to the fundamental weights $\mathbf{\hat{%
\Lambda}}_{0},\mathbf{\hat{\Lambda}}_{1}$\ of $\widehat{su(2)}_{1}$ and can
be written more explicitly as:
\begin{equation}
\chi _{0}^{\widehat{su(2)}_{1}}(\tau )=\sum_{n=0}^{\infty }\left(
2n+1\right) \chi _{2n}^{\mathcal{W}_{2}}(\tau )\;;\;\chi _{1}^{\widehat{su(2)%
}_{1}}(\tau )=\sum_{n=0}^{\infty }\left( 2n+2\right) \chi _{2n+1}^{\mathcal{W%
}_{2}}(\tau )
\end{equation}%
In the $m=2$ case the algebra $\mathcal{W}_{2}$ has central charge $c=1$
and its degenerate representations coincide with those of the Virasoro
algebra. More precisely the primary fields of $ \mathcal{W}_{2}$ with
conformal dimension $L_{0}=k^{2}/4$ have a null vector at level $n+1$ of
the relative representation for any $k\in N$. The characters of these
Virasoro degenerate primary fields are then:
\begin{equation}
\chi _{k}^{\mathcal{W}_{2}}(\tau )=\frac{q^{\frac{k^{2}}{4}}(1-q^{k+1})}{%
\eta (q)}
\end{equation}
Thus the $\mathbf{\hat{\Lambda}}_{0},\mathbf{\hat{\Lambda}}_{1}$\ h.w.
representations of $\widehat{su(2)}_{1}$ are built respectively by the even
and odd degenerate representations of $\mathcal{W}_{2}$.

Now we will derive the identification of the neutral component of the P-A
untwisted sector of our TM and show that it is related to the $\mathcal{W}%
_{m}$ minimal models.

There is just one extended primary field in the neutral component of the P-A
sector of our TM and it can be written in terms of the $\mathcal{W}_{2}$
characters as:
\begin{equation}
\tilde{N}(\tau )=\chi _{0}^{\widehat{su(2)}_{1}}(\frac{\rho }{2}|\tau )=%
\frac{1}{\sqrt{2}}\sum\limits_{n=0}^{\infty }\chi _{2n}^{su(2)}(\frac{\rho }{%
2})\chi _{2n}^{\mathcal{W}_{2}}(\tau )  \label{w2-1}
\end{equation}%
with $\chi _{2n}^{su(2)}(\frac{\rho }{2})=(-1)^{n}.$ Eq.$\left( \ref{w2-1}%
\right) $ tells us that the extended primary field of this sector is defined
by collecting all the $2n$ degenerate primary fields of $\mathcal{W}_{2}$,
everyone with multiplicity one times a residual parity symmetry $(-1)^{n}$.
This extra parity can be seen as the residual symmetry of the finite
$su(2)$ algebra due to cocycles. As a consequence the P-A sector of the TM
contains only degenerate representations of $\mathcal{W}_{2}$. Therefore we
can consider it as a\textquotedblleft minimal sector\textquotedblright\ of
the full theory. Nevertheless in this sector the extended symmetry is given
by $u(1){\times}\mathcal{W}_{2}{{\times} }Z_{2}$. We could take the $Z_{2}$ even
superposition of $\mathcal{W}_{2}$ degenerate characters as it was done in
\cite{Cappelli} but we can see that only the odd superposition given by the
characters $\tilde{\chi}_{(l,b)}$ of the TM belong to a RCFT. It is
necessary to give a well defined $S$ transformation for the characters of
the P-A sector which produces the twisted A-P sector of the TM. In this
sector the $\mathcal{W}_{2}$ characters are non minimal generic
representations.

For generic $m$ the extended characters in the P-A sector contains only $%
\mathcal{W}_{m}$ representations with $\lambda _{a}-\lambda _{b}\in mZ$
with the same structure of the $m=2$ case. Thus only the weights in the
$\mathbf{\hat{\Lambda}}_{0}$ representation contribute and furthermore, by
the isomorphism $\widehat{su(m)}_{1}=su(m){{ {\times} }}\mathcal{W}_{m}$, we have
that in any $su(m)$ multiplet just one is selected out. Instead in the A-P
sector the differences $\lambda_{a}-\lambda _{b}\notin Z$ are not integer
in general.

\section{Conclusions\label{conclusions}}

In this paper, by using the m-reduction procedure, we constructed the
twisted model (TM) for the Jain series at the filling
$\nu=\frac{m}{2pm+1}$, generalizing to the torus topology a recent
construction for the plane\cite{cgm1}. The extended chiral primary fields
were derived both for the twisted and untwisted sectors together with their
corresponding characters. Also their factorization in terms of charged and
neutral components was given. By constructing the complete diagonal
partition function, which turns out to be invariant under $\Gamma_{0}(2)$
it has been shown that the multi-components theory coincides with its P-P
sector content. A detailed relation between the TM, the multi-component and
the $\mathcal{W}_m$ minimal model has been worked out throughout the paper.
For $m=2$ it has been explicitly shown that the P-A sector contains only
degenerate representations of ${\cal W}_2$, so resulting into a ``minimal
sector" of the full theory. Finally it has been shown that the presence of
the A-P sector assures the closure of the TM partition function under
modular transformations. The special $m=2$ case has been also analyzed. In
the appendices it has been reported on the properties of the
$\Gamma_{0}(2)$ group together with the modular transformations of the
$\widehat{u(1)}_q$ and TM characters.

\appendix

{\large \textbf{Appendix A}}

\label{gamma2}\textbf{The }$\Gamma _{0}(2)$\textbf{\ group}

The $\Gamma _{0}(2)$ group is a subgroup of the modular group
$SL(2,Z)/Z_{2}$ generated by $T^{2}$ and $T^{2}ST^{2}$. A matrix
representation of the generators $T^{2}$ and $S$ of $SL(2,Z)/Z_{2}$ is:
\begin{equation}
T^{2}=\left(
\begin{array}{cc}
1 & 2 \\
0 & 1%
\end{array}%
\right) \,\,,\,\,S=\left(
\begin{array}{cc}
0 & -1 \\
1 & 0%
\end{array}%
\right)
\end{equation}%
$\Gamma _{0}(2)$ is the group of elements defined as:
\begin{equation}
\Gamma _{0}(2)=\left\{ \left(
\begin{array}{cc}
a & b \\
c & d%
\end{array}%
\right) \in
SL(2,Z)/Z_{2}\,\,:\,\,a+d\,,\,\,b+c\,\,even;\,\,a+b\,\,odd\right\}
\end{equation}%
Any element $A$ of$\ \Gamma _{0}(2)$ can be represented as follows:
\begin{equation}
A=\left\{
\begin{array}{c}
T^{2a}~~~~~~~~~~~~~~~~~~~~~~~~~~~~~~~\,\,\,\,\forall \,\,a\in Z \\
\\
S_{(a_{1},b_{1})}{{\times} }S_{(a_{2},b_{2})}{{\times} }...{{\times} }%
S_{(a_{N},b_{N})}\,\left.
\begin{array}{c}
\forall (a_{j},b_{j})\in Z{{\times} }Z\,,\,\forall j\in (1,..,N) \\
\forall N\in \mathbf{N}%
\end{array}%
\right.%
\end{array}%
\right.
\end{equation}%
where $S_{(a,b)}=T^{2a}ST^{2b}$. Thus the characterization given for the
subgroup $\Gamma _{0}(2)$ is a direct consequence of the form of this
matrix, in fact
\begin{equation}
T^{2a}=\left(
\begin{array}{cc}
1 & 2a \\
0 & 1%
\end{array}%
\right)
\end{equation}%
and
\begin{equation}
\prod\limits_{j=1}^{N}S_{(a_{j},b_{j})}=\left\{
\begin{array}{c}
\left(
\begin{array}{cc}
2\alpha +1 & 2\beta \\
2\gamma & 2\delta +1%
\end{array}%
\right) \,\,\,for\,\,N\,\,even\, \\
\,\left(
\begin{array}{cc}
2\alpha & 2\beta +1 \\
2\gamma +1 & 2\delta%
\end{array}%
\right) \,\,\,for\,\,N\,\,odd\,%
\end{array}%
\right.  \label{pro-T-S def2}
\end{equation}%
where $\alpha $, $\beta $, $\gamma $, $\delta $ are integers and depend on $%
(a_{j},b_{j})$ and $N$.

\appendix {\large \textbf{Appendix B}}

\textbf{The ECFT of the Laughlin filling}

Here we want to construct the ECFT that describes the Laughlin filling $\nu
=\frac{1}{q}$ with $q=2pm+1$ odd.

This ECFT has to be a free bosonic theory compactified on a circle with
radius $R^{2}=q$, the topological order, that is the degeneracy of the
quantum Hall ground state on the torus, has to coincide with the number of
the extended primary fields of the ECFT\cite{Cristofano}. Furthermore
because the ground state is described by electrons (fermions), the
behaviour required under modular transformations is not the closure under
the entire modular group $SL(2,Z)/Z_{2}$, but only under the subgroup
$\Gamma _{0}(2)$.

The \textquotedblleft smaller\textquotedblright\ RCFT of a free boson with $%
R^{2}=q$(here it means $R^{2}=\frac{2q}{2}$ o by duality $R^{2}=\frac{4}{q}$
) with $q$ odd, is a $\widehat{u(1)}_{k}$ of level $k=2q$. The extended
primary fields are defined by the vertex operator:
\begin{equation}
V_{l}^{\left( 2q\right) }(z)=:e^{i\alpha _{l}\phi \left( z\right) }:\text{ \
\ with }\alpha _{l}=\frac{l}{\sqrt{4q}},\text{ \ \ \ }l=0,.,4q-1
\end{equation}%
and the correspondent characters are:
\begin{equation}
K_{l}^{\left( 2q\right) }(w|\tau )=\frac{e^{-q\pi \frac{\left( Imw\right)
^{2}}{Im\tau }}}{\eta (\tau )}\Theta \left[
\begin{array}{c}
\frac{l}{4q} \\
0%
\end{array}%
\right] \left( 2qw|4q\tau \right)  \label{charge}
\end{equation}%
There are $4q$ extended primary fields and the correspondent characters are
closed under the entire modular group, $SL(2,Z)/Z_{2}$. Thus the ordinary
RCFT $\widehat{u(1)}_{2q}$ does not satisfy the requests of an ECFT for the
Laughlin filling $\nu =1/q$.

The ECFT which does is not an ordinary RCFT. It can be built starting from
$\widehat{u(1)}_{2q},$ by defining in a different way the extended primary
fields. The chiral extended algebra is defined by assembling in any
extended primary field all the primary fields of $\widehat{u(1)}_{2q},$
whose conformal dimensions differ by half integer. We define this extended
theory as $\widehat{u(1)}_{q}$. Its extended primary fields are defined by
the vertex operators\footnote{ These are just the primary fields
representative of the entire conformal blocks that define the extended
primary fields.}
\begin{equation}
V_{a}^{\left( q\right) }(z)=:e^{i\alpha _{a}\phi \left( z\right) }:\text{ \
\ with }\alpha _{a}=\frac{a}{\sqrt{q}}
\end{equation}%
where $a=0,.,q-1$ and the correspondent characters are:
\begin{equation}
\bar{K}_{a}^{\left( q\right) }(w|\tau )=\frac{e^{-q\pi \frac{\left(
Imw\right) ^{2}}{Im\tau }}}{\eta (\tau )}\Theta \left[
\begin{array}{c}
\frac{a}{q} \\
0%
\end{array}%
\right] \left( qw|q\tau \right)
\end{equation}%
There are $q$ extended primary fields and the relative characters are closed
only under the subgroup $\Gamma _{0}(2)$. Thus the $\Gamma _{0}(2)$ RCFT $%
\widehat{u(1)}_{q}$ satisfies the requests of an ECFT for the Laughlin
filling, $\nu =1/q$ .

Finally we notice that the extended theory $\widehat{u(1)}_{q}$ coincides
with the $\Gamma _{0}(2)$-invariant part of the ordinary RCFT $\widehat{u(1)}%
_{2q}$. This is a consequence of the relation among the characters of the
two theories:
\begin{equation}
\bar{K}_{a}^{\left( q\right) }(w|\tau )=K_{2a}^{\left( 2q\right) }(w|\tau
)+K_{2(q+a)}^{\left( 2q\right) }(w|\tau )\ \ \forall a\in \left\{
1,.,q\right\}  \label{pro}
\end{equation}

\textbf{The modular transformations for the }$\widehat{u(1)}_{q}$ \textbf{%
characters}

The modular transformations for the $\widehat{u(1)}_{q}$ characters are
derived by the well known modular transformations for the characters of the
ordinary RCFT $\widehat{u(1)}_{2q}$ appearing in eq.$\left( \ref{pro}\right)
$.

The transformation $T^{2}$ acts as:
\begin{equation}
\bar{K}_{s}^{\left( q\right) }(w|\tau +2)\text{{}}=\text{{}}e^{i4\pi \left(
\frac{s^{2}}{2q}-\frac{1}{24}\right) }\bar{K}_{s}^{\left( q\right) }(w|\tau )
\end{equation}

The transformation $S$ acts as:
\begin{equation}
\bar{K}_{s}^{\left( q\right) }(\frac{w}{\tau }|-\frac{1}{\tau })\text{{}}=%
\text{{}}\frac{e^{iq\pi Re\frac{w^{2}}{\tau }}}{\sqrt{q}}\sum_{s^{\prime
}=0}^{q-1}e^{\frac{2i\pi s^{\prime }s}{q}}\bar{K}_{s^{\prime }}^{\left(
q\right) }(w|\tau )
\end{equation}

\appendix {\large \textbf{Appendix C}}

\textbf{The modular transformations of the TM characters}

\textbf{The case }$m>2$ \textbf{and prime}

\textbf{The modular transformations of the characters of} $\widehat{u(1)}%
_{K}^{\otimes m}$

As we have seen the theory $\widehat{u(1)}_{K}^{\otimes m}$ coincides with
the P-P sector of TM. Thus we write the modular transformation of the
relative characters.

Under $T^{2}$ we get:
\begin{equation}
\tilde{\chi}_{b}^{\mathbf{MC}}(w|\tau +2)=e^{i4\pi \left( \frac{mb^{2}}{%
2(2pm+1)}-\frac{m}{24}\right) }\tilde{\chi}_{b}^{\mathbf{MC}}(w|\tau )
\label{tra per-per T^2}
\end{equation}%
Under $S$ we get:
\begin{equation}
\tilde{\chi}_{b}^{\mathbf{MC}}(\frac{w}{\tau }|-\frac{1}{\tau })=\frac{%
e^{im(2pm+1)\pi Re\frac{w^{2}}{\tau }}}{\sqrt{2pm+1}}\sum_{b^{\prime
}=0}^{2pm}e^{\frac{2i\pi mbb^{\prime }}{2pm+1}}\tilde{\chi}_{b^{\prime }}^{%
\mathbf{MC}}(w|\tau )  \label{tra per-per S}
\end{equation}

\textbf{The modular transformations of the characters built by the }$m$-%
\textbf{reduction procedure }

Here we write the modular transformations for the TM characters $\chi _{(s,%
\tilde{f})}(w|\tau ),\tilde{\chi}_{b}(w|\tau )$.

Under $T^{2}$ we obtain:
\begin{eqnarray}
\chi _{(s,\tilde{f})}(w|\tau +2) &=&e^{\frac{4\pi i}{m}(\frac{s^{2}}{2(2pm+1)%
}+\frac{\tilde{f}}{2}-\frac{1}{24})}\chi _{(s,\tilde{f})}(w|\tau ) \\
\tilde{\chi}_{b}(w|\tau +2) &=&e^{i4\pi \left( \frac{mb^{2}}{2(2pm+1)}-\frac{%
m}{24}\right) }\tilde{\chi}_{b}(w|\tau )
\end{eqnarray}%
Under $S$ it is:
\begin{eqnarray}
\chi _{(s,\tilde{f})}(\frac{w}{\tau }|-\frac{1}{\tau }) &=&e^{i\pi m(2pm+1)Re%
\frac{w^{2}}{\tau }}\sum_{s^{\prime }=0}^{2pm}\sum_{f=0}^{m-1}\mathcal{S}%
_{(s,\tilde{f})}^{(s^{\prime },f)}\chi _{(s^{\prime },f)}(w|\tau ) \\
&&+\frac{e^{i\pi m(2pm+1)Re\frac{w^{2}}{\tau }}}{\sqrt{m\left( 2pm+1\right) }%
}\sum_{b=0}^{2pm}e^{\frac{2i\pi sb}{2pm+1}}\tilde{\chi}_{b}(w|\tau )  \notag
\\
\tilde{\chi}_{b}(\frac{w}{\tau }|-\frac{1}{\tau }) &=&\frac{e^{i\pi
m(2pm+1)Re\frac{w^{2}}{\tau }}}{\sqrt{m\left( 2pm+1\right) }}%
\sum_{s=0}^{2pm}\sum_{f=0}^{m-1}e^{\frac{2i\pi bs}{2pm+1}}\chi
_{(s,f)}(w|\tau )
\end{eqnarray}%
with $\mathcal{S}_{(s,\tilde{f})}^{(s^{\prime },f)}=\frac{1}{m}%
\sum_{j=1}^{m-1}\left( e^{\frac{2\pi i}{m}\left( 2j^{\ast }\right) (\frac{%
s^{\prime 2}}{2(2pm+1)}+\frac{f}{2}-\frac{1}{24})}\left( \mathcal{A}%
_{(m,p,2j)}\right) _{(s,s^{\prime })}e^{-\frac{2\pi i}{m}\left( 2j\right) (%
\frac{s^{2}}{2(2pm+1)}+\frac{\tilde{f}}{2}-\frac{1}{24})}\right) $ where $%
\left( \mathcal{A}_{(m,p,2j)}\right) _{(s,s^{\prime })}$ are the entries of
the $\left( 2pm+1\right) {{\times} }\left( 2pm+1\right) $ matrix $\mathcal{A}%
_{(m,p,2j)}$ that represent the action of the modular transformation $%
A_{(m,2j)}\in \Gamma _{0}(2)$ on the characters of the $\widehat{u(1)}%
_{2pm+1}$ theory. The matrix $A_{(m,2j)}\in \Gamma _{0}(2)$ and the integer
number $j^{\ast }$ are defined in a univocal way by the conditions:
\begin{equation}
\frac{\left( -1/\tau \right) +2j}{m}=\ A_{(m,2j)}(\frac{\tau +2j^{\ast }}{m}%
)\ \ \ \forall j\in (1,..,m-1)  \label{eq.AVII14}
\end{equation}%
In fact eq.(\ref{eq.AVII14}) is satisfied only for:
\begin{equation}
\ A_{(m,2j)}=\left(
\begin{array}{cc}
2j & b-4j\alpha  \\
m & 2j^{\ast }%
\end{array}%
\right) \ ,\ \ \ j^{\ast }=\alpha m-j^{\prime }
\end{equation}%
with $j^{\prime }$ a nonzero integer number with minimal modulo, and $b$ an
odd integer number, defined by the relation:
\begin{equation}
\left( 2j\right) \left( 2j^{\prime }\right) -bm=1  \label{primo}
\end{equation}%
For $m>2$ and prime and for $j=1,...,m-1$, eq.( \ref{primo}) simply
expresses that $m$ and $2j$ are coprime numbers. Finally the value of $%
\alpha $ is uniquely defined by the request $j^{\ast }=1,..,m-1$.

The fact that the matrices $A_{(m,2j)}$ are elements of $\Gamma _{0}(2)$ is
a direct consequence of the characterization given in section (\ref{gamma2}
). By using eq.( \ref{pro-T-S def2}) this matrix can be expressed in terms
of the matrices $T^{2}$ and $S$ by:
\begin{equation}
A_{(m,2j)}=S_{(a_{1},b_{1})}{{\times} }S_{(a_{2},b_{2})}{{\times} }...{{\ {%
{\times}}} }S_{(a_{N},b_{N})}  \label{pro-T-S}
\end{equation}%
where $S_{(a,b)}=T^{2a}ST^{2b}$, are $2{{\times} }2$ matrices, $N$ is an odd
positive integer, $(a_{j},b_{j})\in Z{{\times} }Z$, for $j=1,...,N$.

It is important to point out that the condition $A_{(m,2j)}\in \Gamma
_{0}(2) $ is crucial to obtain the closure under the modular transformation $%
S$ of the characters $\chi _{(s,\tilde{f})}(w|\tau )$, $\tilde{\chi}%
_{b}(w|\tau )$.

{\large \textbf{The case }}$m=2$

The modular transformation of the characters of the P-P untwisted sector
are simply obtained evaluating at $m=2$ eqs.(\ref{tra per-per T^2}) and
(\ref {tra per-per S}).

The modular transformations of the remaining characters are given by ($%
q=4p+1 $):
\begin{eqnarray}
&&\left. T^{2}:\chi _{(s,\tilde{f})}(w|\tau +2)=e^{2\pi i(\frac{s^{2}}{2q}+%
\frac{\tilde{f}}{2}-\frac{1}{24})}\chi _{(s,\tilde{f})}(w|\tau )\right. \\
&&\left. S:\chi _{(s,\tilde{f})}(\frac{w}{\tau }|-\frac{1}{\tau })=\frac{%
e^{2\pi i q Re\frac{w^{2}}{\tau }}}{\sqrt{2q}}\sum_{\mu
=0}^{q-1}\sum_{g=0}^{1}e^{\frac{2\pi i}{2q}(s+q\tilde{f})(\mu +qg)}\tilde{%
\chi}_{(\mu ,g)}(w|\tau )\right.  \label{S-tw}
\end{eqnarray}%
\begin{eqnarray}
&&\left. T:\tilde{\chi}_{(\mu ,g)}(w|\tau +1)=e^{2\pi i\left( \frac{(\mu
+qg)^{2}}{q}-\frac{1}{12}\right) }\tilde{\chi}_{(\mu ,g)}(w|\tau )\right.
\label{S-nontw-1a} \\
&&\left. S:\tilde{\chi}_{(\mu ,g)}(\frac{w}{\tau }|-\frac{1}{\tau })=\frac{
e^{2\pi i q Re\frac{w^{2}}{\tau }}}{\sqrt{2q}}\sum_{s=0}^{q-1}\sum_{\tilde{ f%
}=0}^{1}e^{\frac{2\pi i}{2q}(s+q\tilde{f})(\mu +qg)}\chi _{(s,\tilde{f}%
)}(w|\tau )\right.  \label{S-nontw-1b}
\end{eqnarray}%
Eq.(\ref{S-nontw-1a}) shows that for $m=2$ the characters $\tilde{\chi}%
_{(\mu ,g)}$ are closed under $T$. While in the case $m>2$ and prime number
all the characters of TM are closed only under $T^{2}$; however in any case
the TM is a $\Gamma _{0}(2)$-RCFT.

{\large \textbf{The closure under }}$\Gamma _{0}(2)${\large \textbf{of the
diagonal partition function of TM}}

It is an immediate consequence of the given modular transformations of the
characters of the TM. In order to show more simply the effect of the
modular transformations we define\footnote{ For $m=2$ we have not the terms
$Z_{twist}^{\left( 2\right) +}(w|\tau )$, $ Z_{twist}^{\left( 2\right)
-}(w|\tau )$ and the $S$ transformation becomes
\begin{eqnarray*}
Z_{U(1)_{K}^{\otimes 2}}\left( \frac{w}{\tau }|-\frac{1}{\tau }\right)
&=&Z_{U(1)_{K}^{\otimes 2}}\left( w|\tau \right) \text{ \ ; \ }%
Z_{untwist}^{\left( 2\right) }\left( \frac{w}{\tau }|-\frac{1}{\tau }\right)
=Z_{twist}^{\left( 2\right) }(w|\tau ) \\
&&\left. Z_{twist}^{\left( 2\right) }\left( \frac{w}{\tau }|-\frac{1}{\tau }%
\right) =Z_{untwist}^{\left( 2\right) }(w|\tau )\right. \text{ \ }
\end{eqnarray*}%
}:
\begin{equation}
Z_{twist}^{\left( m\right) +}(w|\tau )=\frac{1}{m}\sum_{s=0}^{2pm}
\left\vert \bar{K}_{s}^{\left( 1+2pm\right) }(w|\frac{\tau }{m})\right\vert
^{2}~~~~~Z_{twist}^{\left( m\right) -}(w|\tau )=\frac{1}{m}%
\sum_{h=1}^{m-1}\sum_{s=0}^{2pm}\left\vert \bar{K}_{s}^{\left( 1+2pm\right)
}(w|\frac{\tau +2h}{m})\right\vert ^{2}
\end{equation}%
In terms of them $Z_{twist}^{\left( m\right) }(w|\tau )$ is written as:
\begin{equation}
Z_{twist}^{\left( m\right) }(w|\tau )=Z_{twist}^{\left( m\right) +}(w|\tau
)+Z_{twist}^{\left( m\right) -}(w|\tau )
\end{equation}

\textbf{The modular transformation }$T^{2}$:
\begin{eqnarray}
Z_{\widehat{u(1)}_{K}^{\otimes m}}\left( w|\tau +2\right) &=&Z_{\widehat{u(1)%
}_{K}^{\otimes m}}\left( w|\tau \right) \text{ \ ; \ }Z_{twist}^{\left(
m\right) }(w|\tau +2)=Z_{twist}^{\left( m\right) }(w|\tau )  \notag \\
&&\left. Z_{untwist}^{\left( m\right) }(w|\tau +2)=Z_{untwist}^{\left(
m\right) }(w|\tau )\right.
\end{eqnarray}

\textbf{The modular transformation }$S$:
\begin{eqnarray}
Z_{\widehat{u(1)}_{K}^{\otimes m}}\left( \frac{w}{\tau }|-\frac{1}{\tau }%
\right) &=&Z_{\widehat{u(1)}_{K}^{\otimes m}}\left( w|\tau \right) \text{ \
; \ }Z_{untwist}^{\left( m\right) }\left( \frac{w}{\tau }|-\frac{1}{\tau }%
\right) =Z_{twist}^{\left( m\right) +}(w|\tau )  \notag \\
Z_{twist}^{\left( m\right) +}\left( \frac{w}{\tau }|-\frac{1}{\tau }\right)
&=&Z_{untwist}^{\left( m\right) }(w|\tau )\text{ \ ; \ }Z_{twist}^{\left(
m\right) -}\left( \frac{w}{\tau }|-\frac{1}{\tau }\right) =Z_{twist}^{\left(
m\right) -}(w|\tau )
\end{eqnarray}

\end{document}